\pdfoutput=1 
\documentclass[sigconf]{acmart}

\usepackage{CJKutf8}
\usepackage[english]{babel}
\usepackage{blindtext}
\usepackage{times}
\usepackage{amsmath}
\usepackage{graphicx}
\usepackage{graphics}
\usepackage{epsfig}
\usepackage{latexsym}
\usepackage{amsfonts}
\usepackage{paralist}
\usepackage{xspace}
\usepackage{mathrsfs}
\usepackage{color}
\usepackage{algorithm}
\usepackage{algorithmicx}
\usepackage{algpseudocode}
\usepackage{listings}
\usepackage{multirow}
\usepackage{url}
\usepackage{tabularx}
\usepackage{subfigure}
\usepackage{subfloat}
\usepackage{etoolbox}

\newcommand{\egg}[1] {}

\newcommand{\smalltitle}[1] {\vspace{6pt} \noindent \textbf{#1}}

\def\ie{\textit{i.e.}\xspace}

\def\etc{\textit{etc.}\xspace}
\def\eg{\textit{e.g.}\xspace}

\def\802{IEEE 802.11\xspace}

\renewcommand\footnotetextcopyrightpermission[1]{} 
\setcopyright{none}

\acmDOI{}

\acmISBN{}

\acmConference[]{}
\acmYear{}
\copyrightyear{}

\acmPrice{}

\begin{document}
\title{NP-RDMA: Using Commodity RDMA without Pinning Memory}
\def\name{NP-RDMA\xspace}
\def\fullname{}
\def\sysname{}



\makeatletter
\patchcmd{\authornote}{\g@addto@macro\addresses{\@authornotemark}}{}{}{}
\makeatother
\author{Huijun Shen, Guo Chen$^{\ast}$, Bojie Li, Xingtong Lin, Xingyu Zhang, Xizheng Wang, Amit Geron, Shamir Rabinovitch, Haifeng Lin, Han Ruan, Lijun Li, Jingbin Zhou, Kun Tan}

\authornote{\emph{Corresponding author: Guo Chen} \protect\\
H. Shen, G. Chen, X. Lin, X. Zhang and X. Wang are with Hunan University, Changsha, Hunan, China. E-mail: \{shj, guochen, inperson, zhangxingyu, xzwang\}@hnu.edu.cn \protect\\
B. Li, A. Geron, S. Rabinovitch, H. Lin, H. Ruan, L. Li, J. Zhou, and K. Tan are with Huawei, China. E-mail: bojieli@gmail.com, \{amit.geron, shamir.rabinovitch, haifeng.lin, ruanhan, jerry.lilijun, zhoujingbin, kun.tan\}@huawei.com}


\renewcommand{\shortauthors}{}

\begin{abstract}
Remote Direct Memory Access (RDMA) has been haunted by the need of pinning down memory regions. Pinning limits the memory utilization because it impedes on-demand paging and swapping. It also increases the initialization latency of large memory applications from seconds to minutes. To remove memory pining, existing approaches often require special hardware which supports page fault, and still have inferior performance. We propose \name, which removes memory pinning during memory registration and enables dynamic page fault handling with commodity RDMA NICs. \name does not require NICs to support page fault. Instead, by monitoring local memory paging and swapping with MMU-notifier, combining with IOMMU/SMMU-based address mapping, \name efficiently detects and handles page fault in the software with near-zero additional latency to non-page-fault RDMA verbs. 
We implement an LD\_PRELOAD library (with a modified kernel module), which is fully compatible with existing RDMA applications. Experiments show that \name adds only 0.1$\sim$2~$\mu$s latency under non-page-fault scenarios.
Moreover, \name adds only 3.5$\sim$5.7~$\mu$s and 60~$\mu$s under minor or major page faults, respectively, which is 500x faster than ODP which uses advanced NICs that support page fault.
With non-pinned memory, Spark initialization is 20x faster and the physical memory usage reduces by 86\% with only 5.4\% slowdown.
Enterprise storage can expand to 5x capacity with SSDs while the average latency is only 10\% higher.
To the best of our knowledge, \name is the first efficient and application-transparent software approach to remove memory pinning using commodity RDMA NICs. 
\end{abstract}

\maketitle

\section{Introduction}

\label{sec:background-RDMA}

Remote Direct Memory Access (RDMA) is prevalent in datacenters due to its low latency and high throughput~\cite{RDMAatScale,HPCC,CloudStorageRDMA,Collie}.
However, an RDMA application must ``pin'' the memory region before local and remote accesses, which fixedly allocates physical memory pages for virtual addresses and makes the memory non-swappable.

Pinning has several drawbacks.
First, pinning affects memory utilization. Even if a page is never accessed, it still occupies physical memory, invalidating on-demand paging. Also, pages cannot be swapped out under memory pressure, thus cannot expand the capacity of DRAMs with inexpensive disks. 
Second, pinning 
complicates programming, since applications cannot utilize existing virtual memory abstraction offered by the operating system (OS). 
Many applications therefore maintain small pinned buffer for communication and copy data on both sending and receiving, undermining the efficiency of RDMA.  
Finally, pinning memory during initialization is slow for applications with large memory footprint, which takes several minutes for hundreds GB of memory.

To remove memory pining, existing approaches often require special Network Interface Card (NIC) hardware that supports page fault. Mellanox ODP\footnote{Stands for on-demand paging.}~\cite{ODP-doc,ODP-paper} modifies NICs to 
re-RDMA the data upon page faults. However, it 
needs to wait for a millisecond-level timeout in case of burdening the network, 
making its latency 100x larger in such scenarios~\cite{ODPpitfall,PART}. 
PART~\cite{PART} proposes a clean-slate NIC architecture to support page faults efficiently, but fleet-wide deployment of new NICs takes time. 
Our goal is to develop a pure software approach to non-pinned RDMA memory using commodity RDMA NICs.

In this paper, we propose \name (Non-Pinned RDMA), which removes memory pinning during memory registration and enables efficiently dynamic page fault handling, requiring no page fault support by NIC hardware. \name adopts several key ideas to address the following challenges: 1) To avoid NIC DMA failures, \name maps fault pages to a special pre-allocated \emph{signature} or \emph{black-hole} page, utilizing IOMMU~\cite{IOMMU}/SMMU~\cite{SMMU} as an additional address mapping layer. 2) To detect page faults in software efficiently, \name adds an additional check on the DMAed data content to see if it matches the signature. 3) To handle page fault quickly, \name converts the original one-sided Read/Write verb into two-sided operations, which swaps in the pages on target memory and redo the transmission.




Particularly, NP-RDMA combines an (1) optimistic one-sided approach and a (2) catch-all two-sided approach:
\begin{enumerate}
	\item The optimistic one-sided approach maps page-fault virtual addresses to a special \emph{signature page} that contains magic numbers, while non-fault pages are mapped normally. 
	One-sided Read operation will always success, but we can check the contents to find the fault pages and use the two-sided approach which converts Reads to reverse Writes. For one-sided Writes, we use an auxiliary memory region that maps fault pages to a special \emph{black-hole page}, and issue an auxiliary one-sided Read after the Write to check whether the read data equals to the written data. When the content checking indicates possible page faults, it then triggers the catch-all two-sided approach.
	\item The catch-all two-sided approach converts any Read/Write to reverse Write/Read. 
	 It temporarily pins the local buffer, sends the local and remote addresses via RDMA Send, then the target software temporarily pins its local memory and Writes the data back to the initiator. As an optimization, small messages are directly sent inline using RDMA Send, thus reducing the latency of pinning pages and reverse Read/Write round-trip-time (RTT).
\end{enumerate}
For optimistic large Reads and Writes, the overhead of checking magic numbers and additional Reads would be considerable.
To minimize per-page overhead, we propose a \emph{page versioning} approach for large optimistic transfers.
Each virtual page is associated with a remotely accessible version, which increments by 1 on each swap-in and swap-out.
We issue 3 consecutive one-sided verbs in parallel: Read version, Read/Write data, and Read version again.
If the versions agree and have odd parity, the page keeps in memory during the data Read/Write.

We utilize x86 IOMMU and ARM SMMU to mobilize the virtual-to-physical address mapping without dynamic modifications to the NIC's Memory Translation Table (MTT), because not all NICs support mutable MTTs and flushing the MTT cache inside NIC is costly.
Synchronizing the IOMMU/SMMU mappings with MMU is not an easy task.
Linux provides the MMU notifiers mechanism so that the \name callback can be invoked when the OS swaps out an page.
However, there is no callback when a page is swapped in.
So, swap-in updates are done lazily when the two-sided approach finds a page fault.

To mitigate the head-of-line blocking inefficiency when a page-fault access is followed by non-page-fault accesses, we enable applications to specify configurable ordering among operations.
Operations with non-overlapping memory and relaxed ordering can be executed out-of-order, without waiting for preceding slow operations blocked by page faults.

We implement \name as an LD\_PRELOAD library for RDMA verbs and a modified kernel module for Mellanox NICs.
Unmodified applications can use NP-RDMA with commodity NICs, e.g., Mellanox ConnectX-4/5/6~\cite{MLX-CX-NICs,MellanoxRPM}.

Experiments show that \name adds only 0.1$\sim$2~$\mu$s latency under non-page-fault scenarios, adds 3.5$\sim$5.7$\mu$s under minor page faults, and adds 60~$\mu$s under major page faults.
We evaluate \name on two end-to-end applications.
Spark initialization time reduces from 120 to 6 seconds with 300 GB memory pool, and TPC-DS benchmark saves 86\% physical memory with only 5.4\% performance slowdown.
By enabling enterprise storage to use SSDs as an expansion of DDR memory, the capacity increases to 5 times while the average end-to-end I/O latency is only 10\% higher.

Our main contributions are summarized as follows.
\begin{itemize}
	\item We propose NP-RDMA, the first efficient and application transparent software approach to remove memory pinning using commodity RDMA NICs.
	\item We propose an one-sided optimistic way to detect fault pages without actually triggering page faults, which skillfully maps fault pages to special signature and back-hole pages, and tracks per-page versions.
	\item We propose a two-sided approach to resolve page faults efficiently which utilizes IOMMU/SMMU to manage memory mapping instead of MTT in RDMA NICs.
	\item Testbed evaluations and enterprise application deployment results show that \name achieves low overhead for non-page-fault remote memory accesses, therefore reducing memory registration latency during initialization and enabling dynamic and swappable memory for RDMA applications.
\end{itemize}

\section{Background and Related Work}
\label{sec:background}
\subsection{RDMA and the Memory Pinning}

\subsubsection{RDMA Background}
RDMA has been widely deployed in data centers to save significant CPU resources, reducing the average and tail latency, and improve the overall throughput. By offloading all the network stack processing 
into the RDMA NIC (RNIC), it provides the ability to bypass the remote CPU and OS to access remote memory directly. RDMA applications post work requests (WRs) to RNICs through verbs API. Read/Write verbs can directly read/write remote memory, without involving the remote CPU (called one-sided verbs). Traditional message-passing verbs (\ie, Send/Recv) are also supported, which require explicit participant of the remote CPU in data transmission (called two-sided verbs).  


In order to access host memory directly using the RNIC (through DMA), RDMA applications
need to register memory regions (MRs) before communication. It involves copying OS page table entries of the corresponding memory to the memory transaction table (MTT) in the RNIC. According to the MTT, RNICs then can translate the application-wise virtual address (VA) into the host-wise physical address (PA), bypassing the CPU. 
In virtualized scenarios, there is another address translation between the real PA and the ``PA'' viewed by IO devices like RNICs (called IOVA). X86 IOMMU~\cite{IOMMU} or ARM SMMU~\cite{SMMU} conducts such translation. 
To ensure security, two keys (\texttt{lkey} and \texttt{rkey}) are generated for each MR for local and remote access. Memory in an MR can only be accessed with the correct VA associated with the corresponding access keys.


\subsubsection{Memory Pinning in RDMA}
Pinning roots from the current virtual memory architecture. To provide each process the abstraction of exclusively holding a large and linear memory address space, the mapping from VA to PA can be dynamically changed by the OS. Memory pages can be swapped out (paging) to disk if the memory spare space is limited, or even will not be allocated until the first access. Under such case, the memory mappings in the MTT would be invalid and DMA failures would occur in RNIC. 
Therefore, RDMA will "pin" (or "lock") the page when registering memory, that is, lock the mapping relationship of VA-PA and ensure that this memory area will always exist in physical memory without paging during the communication. 



With the pin-down caches in MTT, RNIC can work without the cooperation of the host and have ultra low-latency and high-throughput in network communication. However, this sacrifices the flexibility of using memory. Obviously, pinning all the memory that may be accessed in RDMA brings very poor memory utilization ~\cite{AnalysisofIBMR,HybridIO,2003DMAreg,Tezuka1998PindownCA,PVFS2003,Levy2020TheCF,Denis2011AHP,Wu2004UnifierUC,Rashti2011ExploitingAB}. Moreover, pinning incurs costly overhead during memory registration. For example, MR registration takes about 120 seconds for 300 GB of memory (see \S\ref{sec:eval-spark}), which is several times higher than the actual data transmission time in RDMA.  
The memory bottleneck has greatly hindered the further development of RDMA, making it a necessary requirement to efficiently support dynamic memory paging.

\label{sec:related-work}
\subsection{Existing Solutions and the Drawbacks}
\label{sec:relate}

\subsubsection{Traditional Memory Management in RDMA Applications}
\label{sec:back-soft-methods}

\begin{figure}
    \centering
    \includegraphics[width=0.4\textwidth]{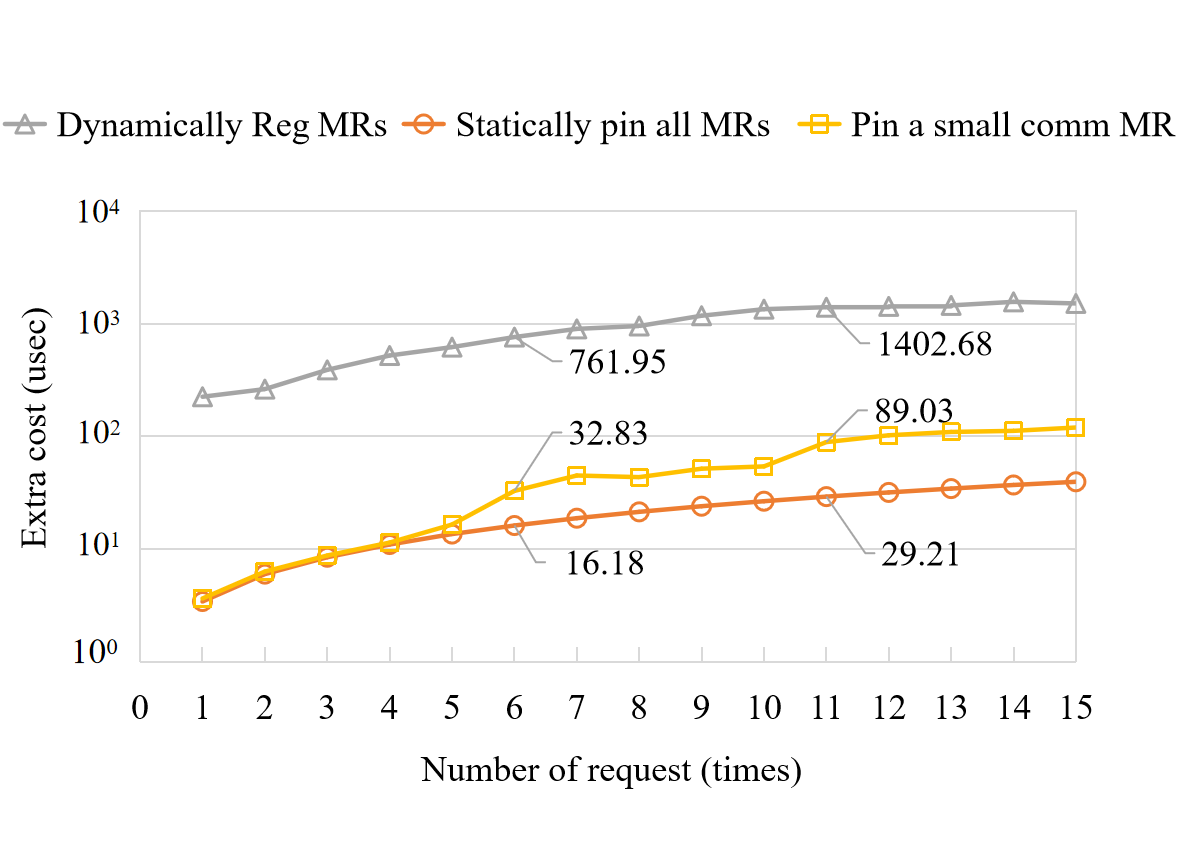}
    \vspace{-10pt}
    \caption{Performance overhead of managing memory in RDMA applications, limited by the memory pinning problem.}
    \label{fig:unwise-apps}
    \vspace{-15pt}
\end{figure}

Applications typically use two approaches to avoid pinning all their memory that will be possibly used in RDMA communication. 
The first is to dynamically register/de-register MRs ~\cite{2003DMAreg,0copyoverIB,nieplocha2002protocols,2009fastMR,Tezuka1998PindownCA,Chen2022UnderstandingRM,Ou2006MRRCA,0copy2022,2021maxRDMAbenefit,2009minRDMAcost,Levy2020TheCF}.
However, each MR registration takes $\sim$50~$\mu$s, making the end-to-end latency one order of magnitude higher.
In addition, for one-sided Read/Write, the application must use two-sided Send/Recv to notify the remote software to perform the MR registration, wasting CPU resources.
The second approach is to pin a small ``communication buffer'' and explicitly copy data to the send buffer and from the receive buffer ~\cite{AnalysisofIBMR,HPRDMA2003,Naos,RDMAinLSM2021,2021maxRDMAbenefit,PVFS2003,2009minRDMAcost,RDMAforKVS2014,2017LITE,mietke2006reducing}. Nevertheless, this data copying approach essentially eliminates the benefits of zero-copy in RDMA.  
We conduct a small experiment to show their performance overhead. We generate 15 Reads, where the size of the 1st-5th Reads is 64B, and the size of the 6th-10th Reads is 128B, and the size of the 11th-15th ones is 192B. Detailed testbed settings are introduced in \S\ref{subsec:testbed-setup}. We emulate three application modes: 1) statically pinning all the buffer (MR) used for Reads, 2) dynamically registering/deregistering the buffer (MR) before/after a Read, and 3) pinning a small dedicated communication buffer (64B) and explicitly splitting large Reads into 64B ones and copying data out. Figure \ref{fig:unwise-apps} shows the results. Such memory management manners limited by the pinning problem can slowdown the RDMA performance by 39\%$\sim$97\%, and dynamic MR registration costs even more. 

\subsubsection{Supporting Page Fault in NICs}
There are a few recent works trying to directly support page faults in RNICs~\cite{ODP-paper,PART}, thus to remove the burden of pinning memory in applications. However, the requirement of new NIC hardware impedes their deployment. Moreover, their performance under page fault is inferior due to the non-optimal collaboration between the NIC hardware and the host OS.    

Mellanox (now NVIDIA) On-Demand Paging (ODP)~\cite{ODP-presentation,ODP-paper,ODP-doc} enables RDMA-triggered page faults on RNICs.  ODP modifies the RNIC such that, when doing an RDMA verb that accesses a VA which is not mapped into a existing PA, the RNIC can ask a driver to raise an interrupt to query for the OS kernel. Then, the OS tries to swap in the VA and then returns the corresponding PA to the RNIC by updating its MTT. 
Then, the initiator RNIC will retransmit the operation through network after a timeout.

\begin{figure}[htbp]
    \centering
    \includegraphics[width=0.45\textwidth]{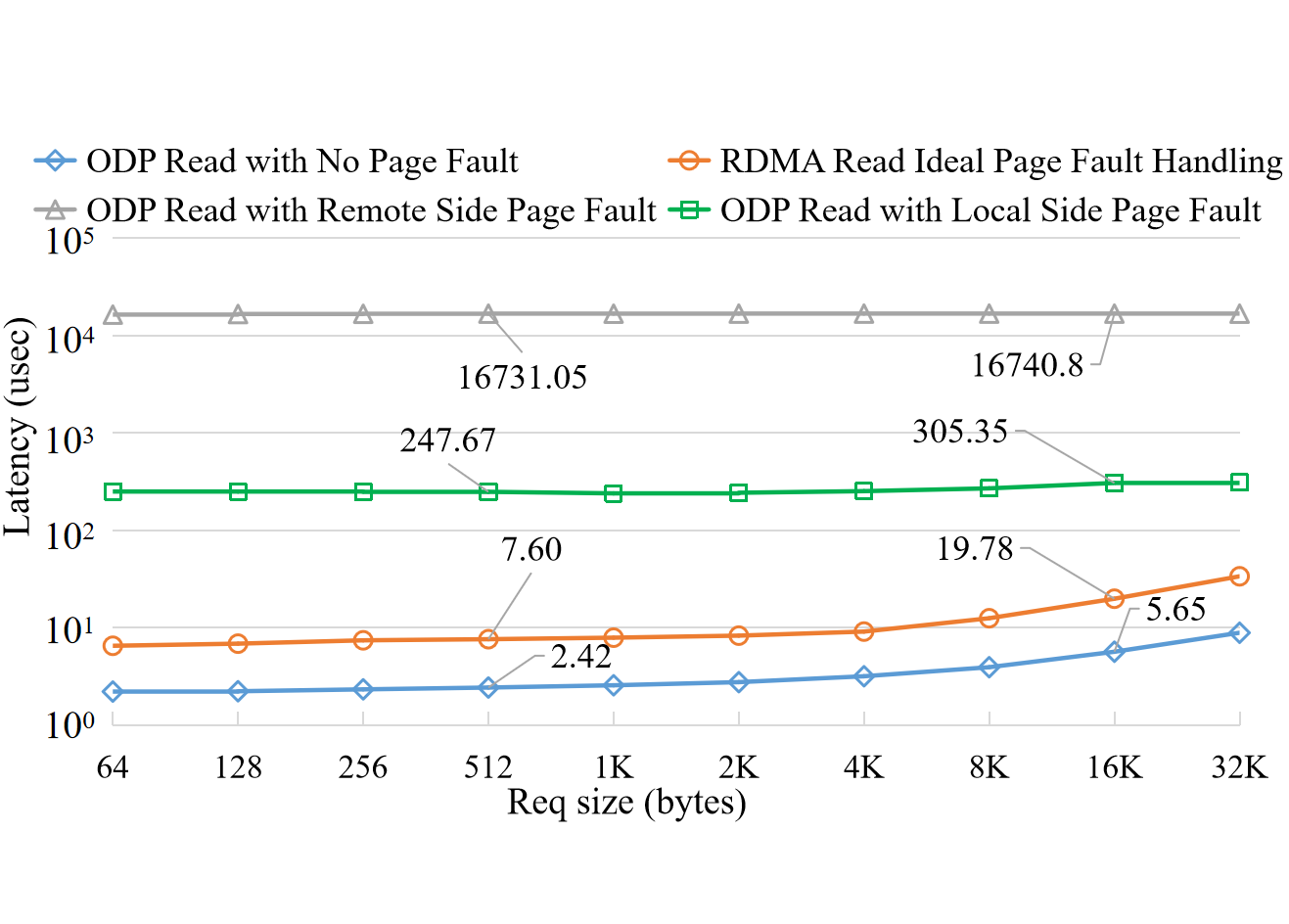}
    \vspace{-15pt}
    \caption{ODP performance under local and remote minor page faults when doing an RDMA Read. The time of RDMA ideal page fault handling is calculated by summarizing the time of doing two normal RDMA Reads and the time of handling one minor page fault in the OS.}
    \label{fig:odp-drawbacks}
    \vspace{-15pt}
\end{figure}

There are three major drawbacks in ODP:
first, communication between the local RNIC and the local host OS is costly upon page fault. Particularly, it takes significant time a) for the RNIC to notify the OS by raising interrupt when page fault happens, and b) for the OS to notify the RNIC and update its MTT when the page has been swapped in. To verify the latency overhead, we deliberately generate a local page fault when an RNIC Reads some remote memory to a local address. Fig.~\ref{fig:odp-drawbacks} shows that (detailed settings in \S\ref{subsec:testbed-setup}), for a local minor page fault which needs only about several $\mu$s to handle in OS, the communication between the RNIC and OS incurs about 28x$\sim$37x extra latency (about 231$\sim$286$\mu$s). 

Second, communication between the local RNIC and the remote host OS/RNIC is even more complicated and costs more. When remote page fault happens (\eg, an RNIC Write to some remote addresses that do not exist in the remote memory), 
	besides the local communication discussed before, 
	the remote side also needs to interact with the local RNIC. Particularly, it needs to notify the local RNIC to pause retransmitting this WR, otherwise the network and the NIC processing capacity will be wasted. Moreover, after the memory has been swapped in, the remote OS needs to quickly notify the local RNIC to resume the operation and retransmit the RDMA WR through network. 
	Such manner significantly complicates the RNIC implementation and the NIC retransmission takes time. For example, to avoid the implementation complexity of getting notification from the remote OS, Mellanox ODP RNICs implements a simple conservative retransmission scheme~\cite{ODPpitfall,PART}. After a remote page fault, the local RNIC will conservatively retransmit the WR after a ms-level timeout. Fig.~\ref{fig:odp-drawbacks} shows that, even for a remote minor page fault, ODP still waits more than 16ms to retransmit the WR, which exaggerates the page fault handling latency by about 496x$\sim$2514x. 
	Moreover, if there are subsequent WRs on the fly, they will all be dropped and retransmitted until the previous WR which encounters page faults has been completed\footnote{A possible better solution may be to selectively retransmitted the page-fault WR and temporarily buffer all the subsequent non-page-fault ones at the remote side. However, implementing high-performance selective transmission on RNICs is also a difficult task~\cite{IRN,MELO}.}. Significant network bandwidth and NIC processing capability will be wasted under such cases.   

Third, the requirement of new NIC hardware increases deployment difficulties. RDMA ODP is still not available in RNICs from some vendors other than Mellanox~\cite{huawei-in200}. Also, Mellanox legacy RNICs such as ConnectX-3~\cite{mellanox-cx3} do not support ODP. Moreover, it is hard to incrementally deploy ODP in a datacenter since it needs the support of the RNICs on both two communication sides. This impedes the deployment.   

More recently, PART~\cite{PART} proposes a new NIC hardware that improves the page fault performance compared to ODP. However, it is a new clean-slate architecture that integrates the RNIC in the same chip or package with the main processor, which is still hard to deploy in existing datacenters. Moreover, it still has some of the aforementioned drawbacks of supporting page-fault in the hardware.

\smalltitle{Wrap-up of existing solutions.} There still lacks a solution which is easy to deploy requiring no hardware modifications, incurs little performance overhead to normal RDMA operations and is transparent to applications, and handles page faults quickly when they happens.   











\subsection{Goals and Challenges of \name}

\subsubsection{Goals} 
Our desire is a pure software approach to non-pinned RDMA which is transparent to existing applications using commodity RDMA NICs, only incurring slight latency or CPU overhead to normal RDMA operations, and handling page fault fast. 

\subsubsection{Challenges} 
Bearing this goal in mind, \name faces several critical challenges: 
\begin{itemize}
	\item The first challenge is without page-fault support in existing RNIC hardware, how to avoid DMA failures when the target VA is not in the memory?
	\item Without notification from the RNIC hardware, how to detect page faults in software efficiently? 
	\item After detecting the page fault in software and swapping-in the fault page, how to quickly notify the local or remote RNIC to redo the WR and (R)DMA the data again?
\end{itemize}

Next, we introduce the design of \name in more details and show how the above challenges are addressed.

\begin{figure*}[htbp]
	\centering
	\subfigure[Read messages workflow.]{
		\label{fig:smsg-read}
		\includegraphics[width=0.32\textwidth]{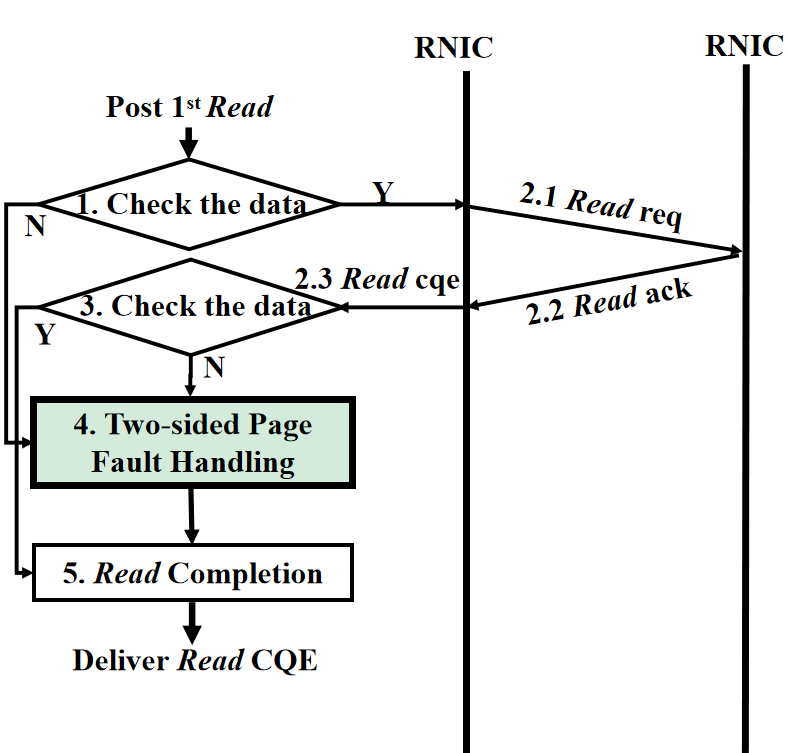}
	}
	\subfigure[Write small messages workflow.]{
		\label{fig:smsg-write}
		\includegraphics[width=0.32\textwidth]{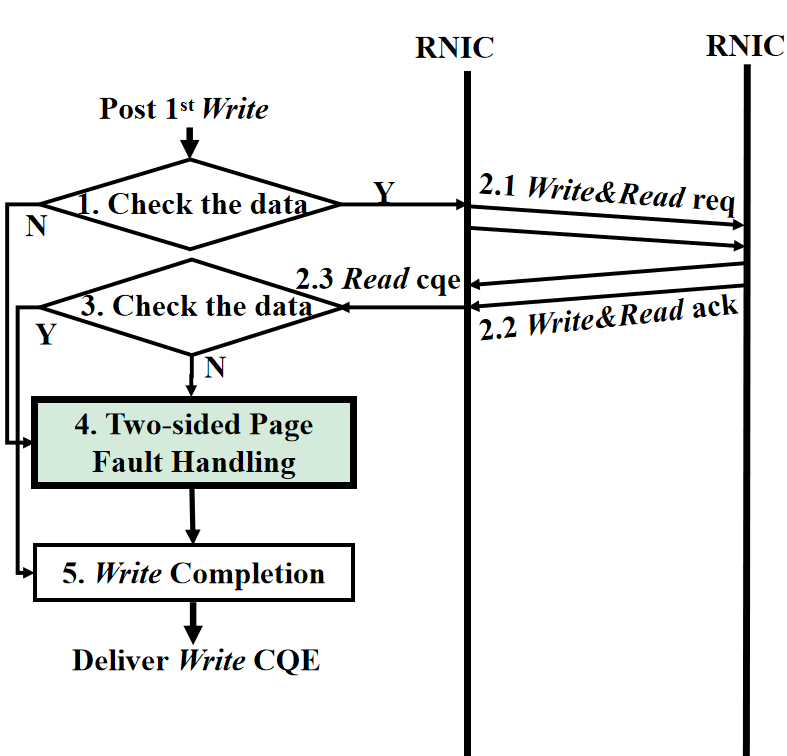}
	}
	\subfigure[Write large messages workflow.]{
		\label{fig:bmsg}
		\includegraphics[width=0.32\textwidth]{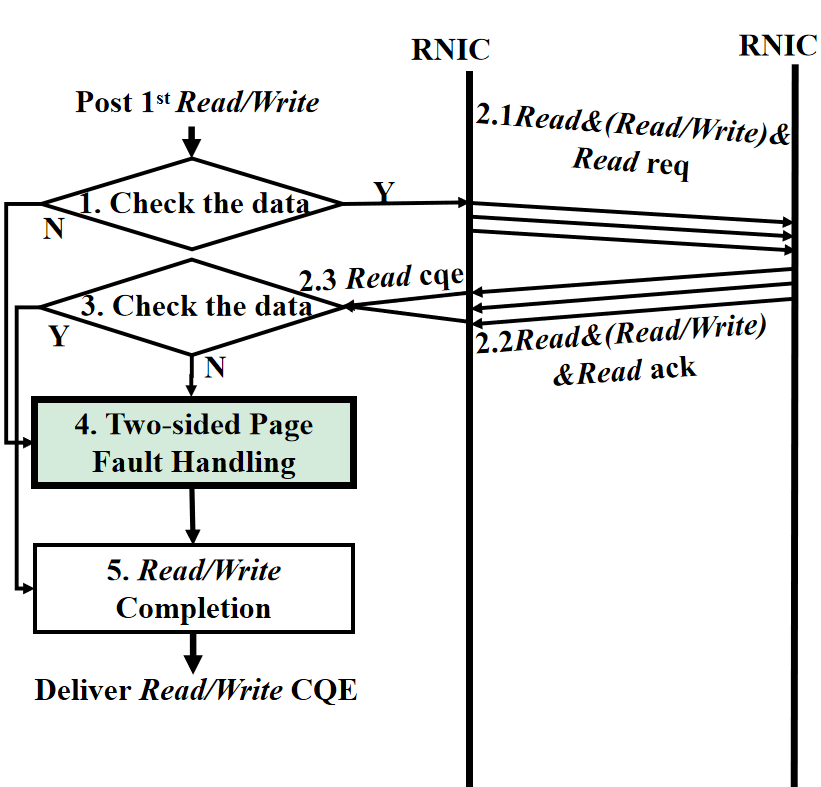}
	}
	\vspace{-15pt}
	\caption{Workflow of optimistic one-sided transfer in NP-RDMA.}
	\label{fig:one-sided-workflow}
	\vspace{-10pt}
\end{figure*}

\section{\name Design}
\label{sec:design}

NP-RDMA first attempts to complete a one-sided RDMA Read/Write in an optimistic way without introducing too much overhead (Sec.\ref{subsec:optimistic}).
If the optimistic one-sided operation detects a page fault, the operation will be converted to a two-sided operation which involves the remote CPU to resolve the page fault (Sec.\ref{subsec:two-sided}).
Sec.\ref{subsec:ordering} extends the RDMA semantics to support configurable RDMA transaction ordering to parallelize page-fault handling and normal operations.


\subsection{Optimistic One-sided Read/Write}
\label{subsec:optimistic}

Fig. \ref{fig:one-sided-workflow} shows the workflow of the one-sided Read/Write in NP-RDMA.
The first step is to check whether the local page is in memory, and fall back to two-sided page fault handling if not.
Second, it issues one-sided Read/Write verbs to perform the actual operation.
For Write, it also issues an auxiliary Read to check whether the Write has succeeded.
To handle large Writes efficiently, Fig.~\ref{fig:bmsg} checks the version of a page rather than its contents.
Third, it checks the data to see whether the Read/Write succeeds, and generates completion if so.
If the check fails, step 4 falls back to the two-sided way.

\subsubsection{Signature Page with Magic Number}
\label{subsec:magic-number}

\begin{figure}[htbp]
    \centering
    \subfigure[Invalid VAs mapped to signature or black-hole pages (for small MRs)]{
    \label{fig:mappings-signature}
    \includegraphics[width=0.22\textwidth]{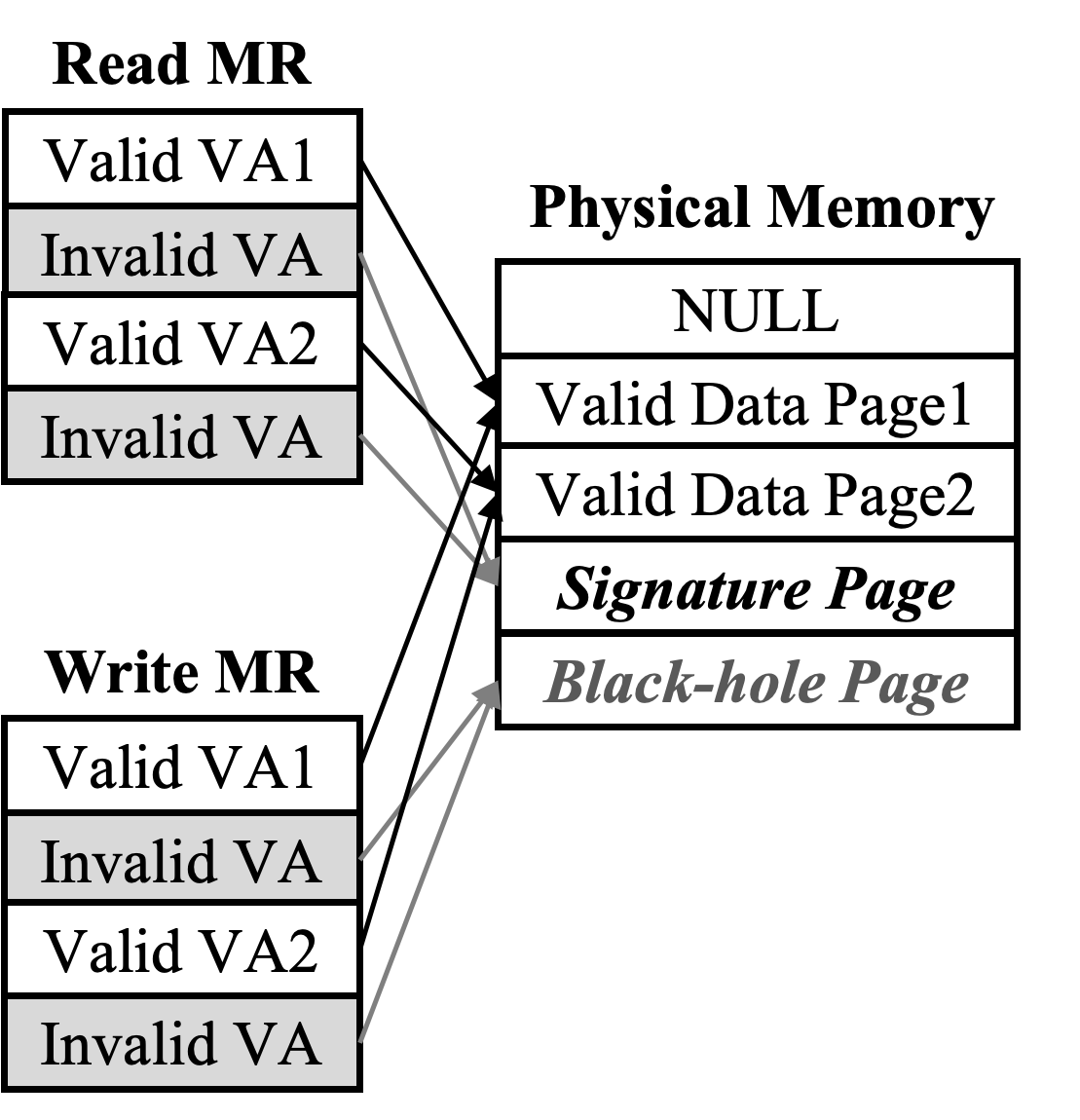}
    }
    \hspace{1pt}
    \subfigure[Invalid VAs mapped to version datasheets or black-hole pages (for large MRs)]{
    \label{fig:mappings-version}
    \includegraphics[width=0.22\textwidth]{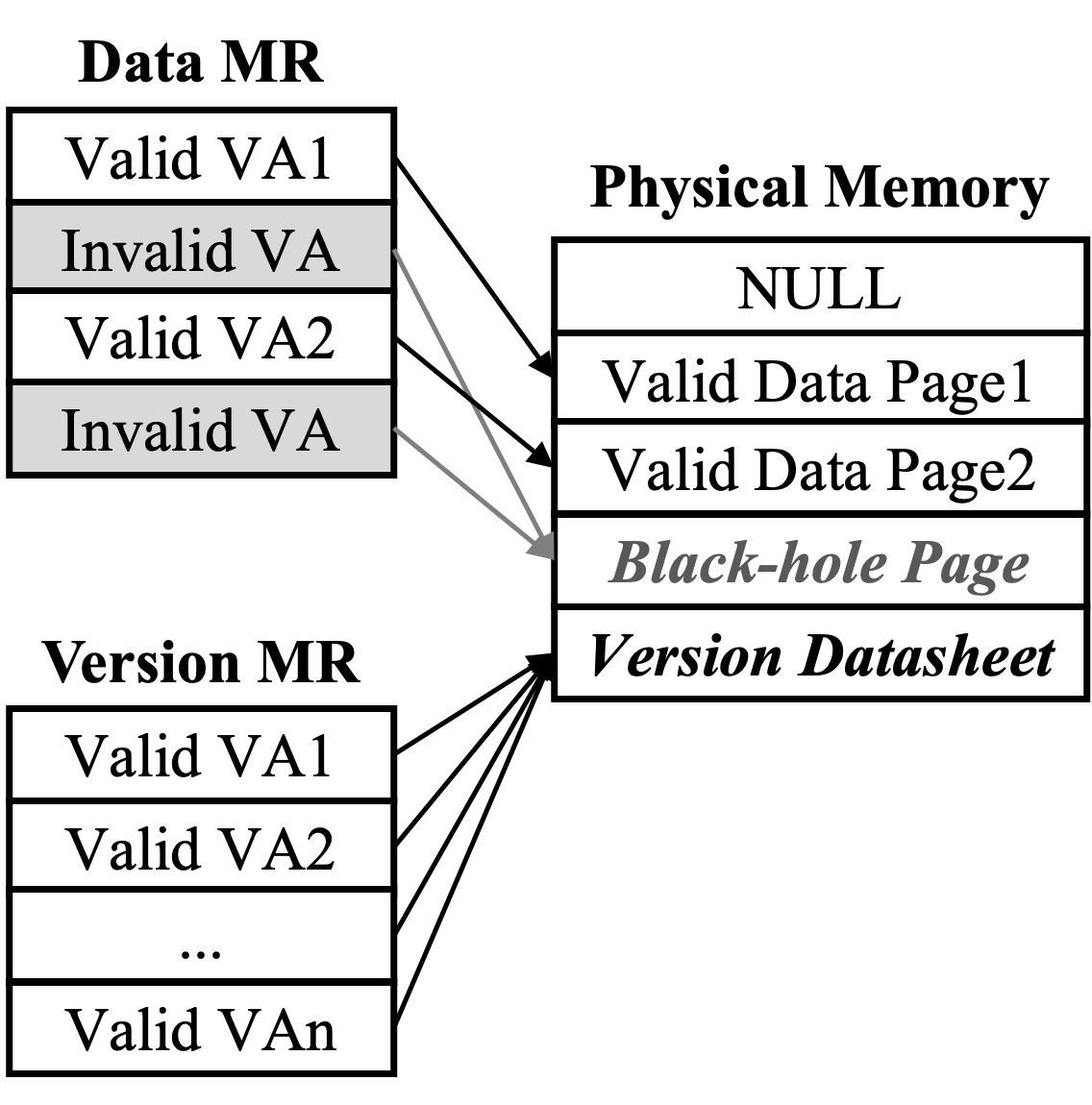}
    }
	\vspace{-15pt}
    \caption{MR mappings in \name.}
    \label{fig:mr-mappings}
    \vspace{-10pt}
\end{figure}


We design a novel mechanism to detect page faults with one-sided verbs without actually triggering page faults.
Rather than mapping non-existing virtual addresses to NULL, when pages are swapped out by the OS, we utilize x86 IOMMU or ARM SMMU to map the IOVAs to a special page called \textit{signature page} filled with magic numbers (e.g., 0xdeadbeef repeated 1K times), as shown in Fig.~\ref{fig:mappings-signature}.
After reading data from the remote side, the initiator compares the data against the magic numbers.
If the data equals to the magic number, it may be the actual data or a page fault.
We do not know which is the case, and always use the two-sided page fault handling approach to redo the transfer.
If the actual data matches the magic number by coincidence, the result would still be correct, but it would incur 2 RTTs of extra latency.

\textbf{Handling local faults.}
The scratchpad approach above can only handle remote Read page faults.
Considering local page faults of RDMA Read and remote page faults of RDMA Write, we need a way to handle DMA write faults.
If we simply enable the MR to handle DMA writes, the magic numbers in the signature page would be overwritten.
To solve this problem, we allocate another special page namely \textit{black-hole page}, and create an auxiliary Write MR which maps non-existing virtual pages to the black-hole page.
As its name indicates, the black-hole page is never read.
The original MR is renamed as Read MR because it is never written by DMA.
RDMA Read uses Write MR on the initiator and Read MR on the target.
To enable the host CPU to read the magic numbers correctly when there are local page faults, we remap the Read MR to a virtual memory space in the process.
If the data equals to the magic number, it may be local or remote page fault, or a coincidence.
Otherwise, the one-sided Read is successful, and the only overhead is checking the read data with host CPU.

\textbf{Handling Writes.}
RDMA Writes are much tougher.
RDMA Write uses Read MR on the initiator and Write MR on the target. 
After the one-sided Write, we do not have any clue about whether the data is written into a normal page or the black-hole page.
So, we issue an additional Read after the original Write, and then check the response.
The Read and Write can be in parallel because RDMA QP is strictly ordered.
The auxiliary Read is written into a temporary local buffer.
If the read response equals the data to be written and it does not match the magic numbers, the Write is successful.
Otherwise, the local or remote buffer has a page fault, or coincidence, or the written data has been altered before the read.

\textbf{Check per DMA granularity.}
An optimization is that we actually do not need to check every byte against the magic number.
Checking 4 bytes per page is insufficient because the NIC does not read a page atomically.
The page may be swapped out in the meantime of NIC access, and the first part of the read data is normal but the remaining part matches the magic numbers.
Checking first and last bytes do not help because the page may be swapped out and swapped in during the Read.
So, the only safe way is to check 4 bytes per NIC DMA granularity.
FaRM~\cite{farm} makes use of the finding that 64-byte RDMA Read and Write are atomic.
Using a PCIe protocol analyzer, we find that for larger DMA transfers, Mellanox ConnectX-5/6 NICs~\cite{mellanox-cx5,mellanox-cx6} issue DMA operations according to the maximum TLP size of PCIe DMA Read and Write, which can be found via \texttt{lspci}.
We only need to check 4 bytes per the minimum atomic DMA Read/Write size in initiator and target, which is typically 256 bytes.
To ensure page swap-out does not occur while a DMA transaction is in-flight, we use MMU notifiers~\cite{mmu-notifier} to flush IOMMU/SMMU when a page is swapped out.
Although the PCIe root complex may split DMA read completions to 128-byte chunks~\cite{kalia2016design}, IOMMU flushing still makes sure that a PCIe Read started before swap-out completes before the page is reclaimed.
Notice that we still read all the data from target to initiator because small work requests have low throughput.

\textbf{Pre-check local page faults.}
In the case of local page faults, the approach above would still introduce one RTT to detect them.
We can check the pages against magic numbers before issuing the optimistic one-sided Read/Write.
The check introduces an overhead of 10~ns per page, but it saves one RTT for a local page fault.

\textbf{Batch unsignaled Writes.}
For Writes, the Write-Read pattern introduces a performance problem because inside the target NIC the Read needs to wait for the Write DMA to complete, adding 1.5~$\mu$s latency and reducing small message throughput by 90\%.
We find that applications issuing consecutive Writes often use \emph{unsignaled} WRs that do not generate CQEs.
The application on initiator knows the Writes are completed when it receives the CQE of a \emph{signaled} WR.
So, we can postpone the Reads to the first \emph{signaled} WR in the same QP, and the Write-Read latency is amortized.

\subsubsection{Page Versioning}
\label{subsec:page-versioning}

Signature page approach clearly incurs significant overhead for large Write operations because it doubles the bandwidth usage.
So for large Writes, We propose a complementary approach to the above one. 
Realizing that we need to know whether the page is always in swap-in state when the target NIC is accessing it, we explicitly maintain a version for each page.
The version is incremented by 1 on both swap-in and swap-out, and the version table is a pinned MR that can be remotely accessible.
When the MR is registered, the version is 1 if the page exists and 0 if not.
So, odd versions indicate existing pages and even versions are fault pages.

To read or write remote data, we first read its version, then perform the actual read or write, and finally read its version again.
If the two versions are equal, then no page swapping occurs between the two read version operations.
So, the actual read or write succeeds if the two versions are equal and have odd parity.
Otherwise, we fall back to the two-sided mechanism to handle page faults.
To detect local page faults, we also need to read and compare the local versions before and after the remote request, which has negligible overhead.
Page versioning adds two small RDMA reads for each actual read or write.
Since the version table is contiguous in memory, and 4-byte versions are sufficient to prevent reusing a version during an RDMA operation, the read overhead is 8 bytes per page, or 0.2\%.
The overheads of the signature page and page versioning approaches will be compared in Sec.\ref{subsec:non-page-fault}.

\subsection{Two-sided Page Fault Handling}
\label{subsec:two-sided}

\begin{figure}[htbp]
    \centering
    \subfigure[Read/Write small messages workflow.]{
    \label{fig:two-sided-smsg}
    \includegraphics[width=0.45\textwidth]{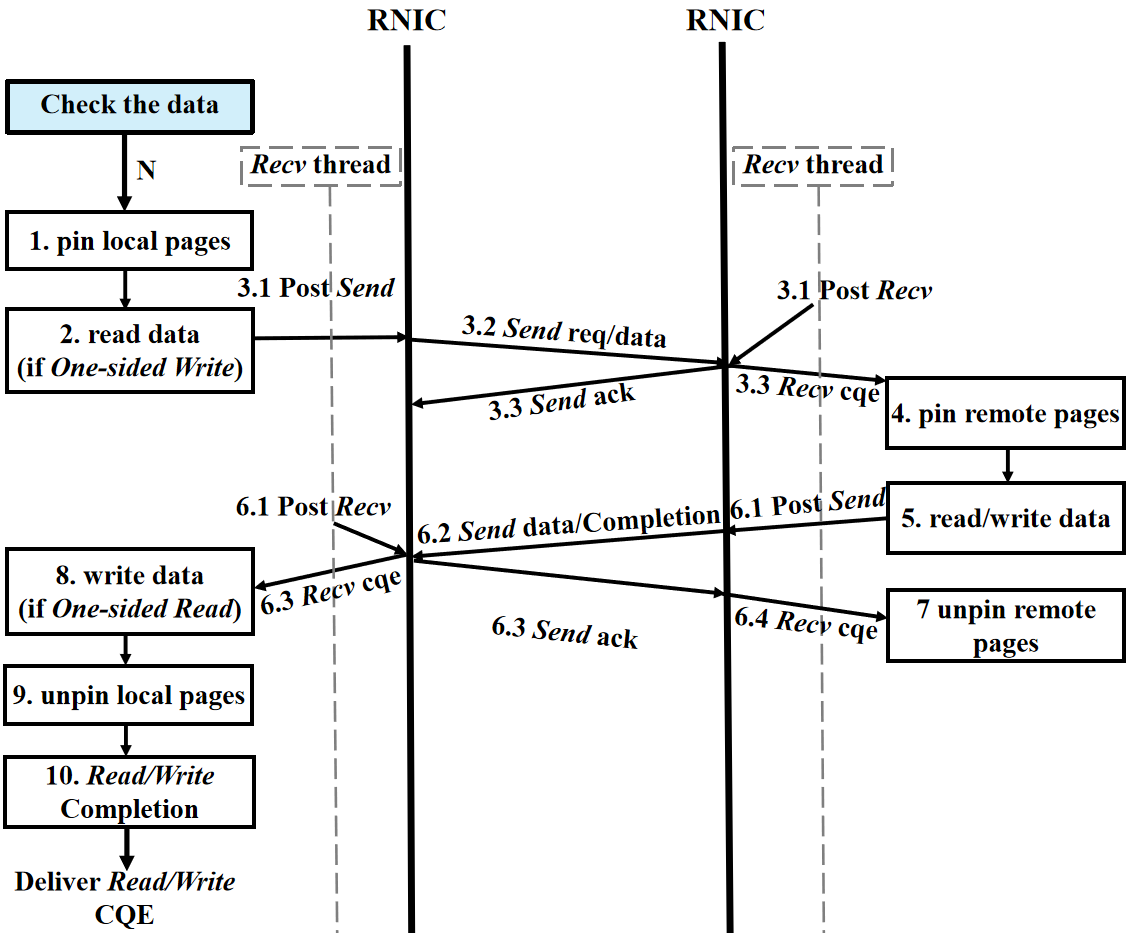}
    }
    \subfigure[Read/Write large messages workflow.]{
    \label{fig:two-sided-bmsg}
    \includegraphics[width=0.45\textwidth]{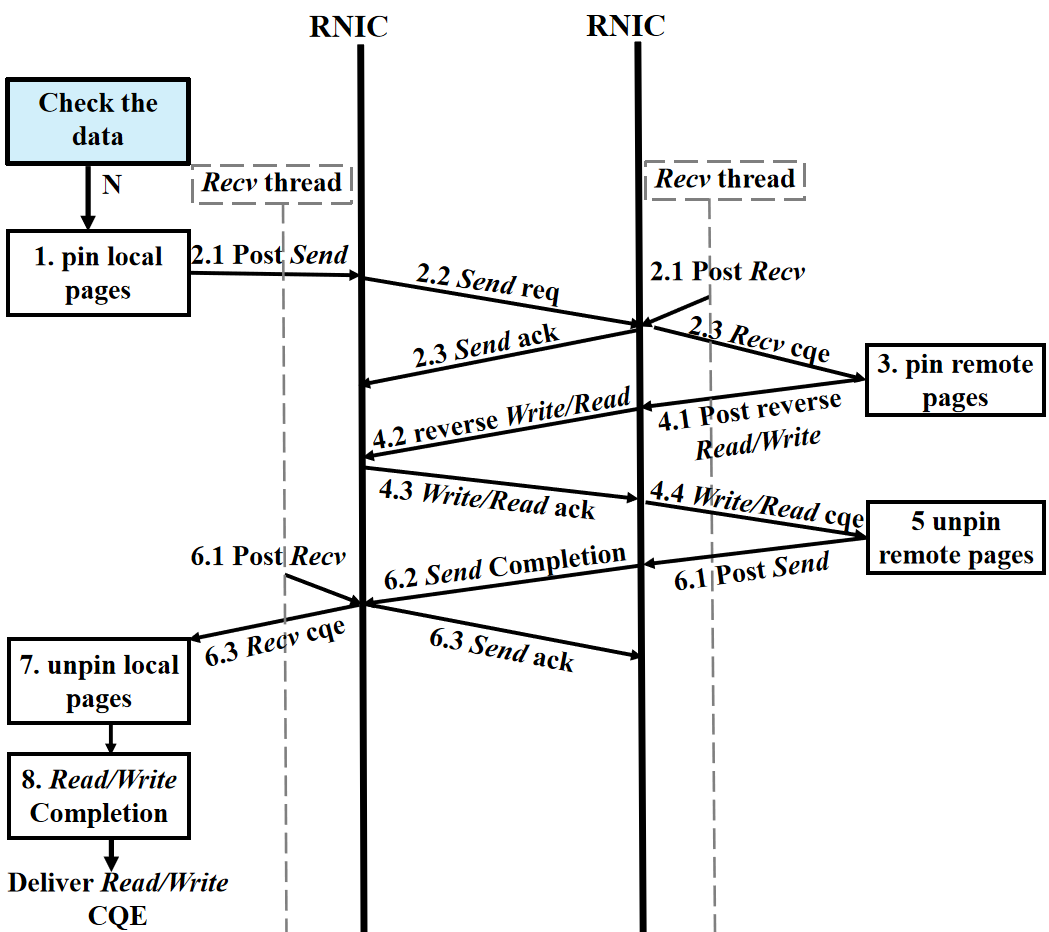}
    }
    \vspace{-15pt}
    \caption{Two-sided basic workflow of NP-RDMA.}
    \label{fig:two-sided-workflow}
    \vspace{-15pt}
\end{figure}


When the optimistic approach detects a possible page fault, we need a way to swap in the pages.
If the page fault is at local buffer, we can simply swap in the page and try again.
However, for remote page faults, we need to notify the software on target host to swap in the pages.
We establish a control QP as the notification channel, which is associated with a small pinned buffer on both initiator and target to make sure that the control channel does not trigger page faults.

The page-fault handling process is shown in Fig.~\ref{fig:two-sided-workflow}, and the major steps are summarized as follows:
1) the initiator pins local buffer and updates IOMMU/SMMU mapping; 2) the initiator sends a request message to the target including Read/Write opcode, local address, remote address and length;
3) the target receives the message with a polling thread, pins the remote buffer, and updates IOMMU/SMMU mapping;
4) the target issues a reverse Write/Read operation as the local and remote buffers are all temporarily pinned;
5) upon completion of the reverse Write/Read, the target unpins the remote buffer and sends a completion message to the initiator;
6) the initiator unpins the local buffer and generates CQE to notify the application.

In steps 3) and 6), the target and initiator need to poll request and completion messages.
To minimize polling overhead, we share a polling thread among all CQs in a process which handles both initiator and target tasks.
So, each process only requires one additional thread.
If its overhead is still too large, the application can configure to use interrupt mode at the cost of additional $\sim$5~$\mu$s latency for page-fault operations. 

\textbf{Short message inline optimization.}
To avoid the additional RTT and the overhead of temporary pinning, we can send the data inline via the pinned control command queue.
This would require copying data from the application's send buffer to the pinned send queue, and from the pinned receive queue to the application's receive buffer.
In our measurements, messages smaller than or equal to 1KB is more efficient to be sent inline.

\subsection{Configurable Ordering}
\label{subsec:ordering}

Readers may have already found ordering problems of the optimistic one-sided approach.
If a write operation fails, the subsequent write may succeed, violating the ordering requirement in an RDMA QP.
Waiting the last request to complete before sending the next would result in poor performance.
The mismatch between varying access latency and strict ordering is a fundamental problem.
Even if we have a new NIC hardware that can handle page faults, a read still needs to wait until the last page-fault read to complete, which takes $\sim$~50$\mu$s to read from the SSD.

Realizing this problem, we aim to provide configurable ordering to the applications.
In most cases, what we care is the ordering between operations with overlapping memory.
The initiator maintains a table to track in-flight operations.
If an operation has overlapping memory with an in-flight operation, it and its subsequent operations in the same QP will be stored in a pending request buffer, waiting for the in-flight operation to complete.
We find that most applications can work well with the overlapping memory heuristic.

In some special cases (e.g., metadata and data), the application also desires strict ordering between work requests.
We provide two ordering bits: \emph{order-before bit} and \emph{order-after bit} in the work request.
When the order-before bit is set, the work request always wait for all in-flight operations to complete, making it strictly ordered with preceding work requests.
When the order-after bit is set, we wait until the work request to complete before issuing new work requests, making it strictly ordered with subsequent work requests.

\section{Implementation}
\label{sec:implementation}

\begin{figure}
	\centering
	\includegraphics[width=0.5\textwidth]{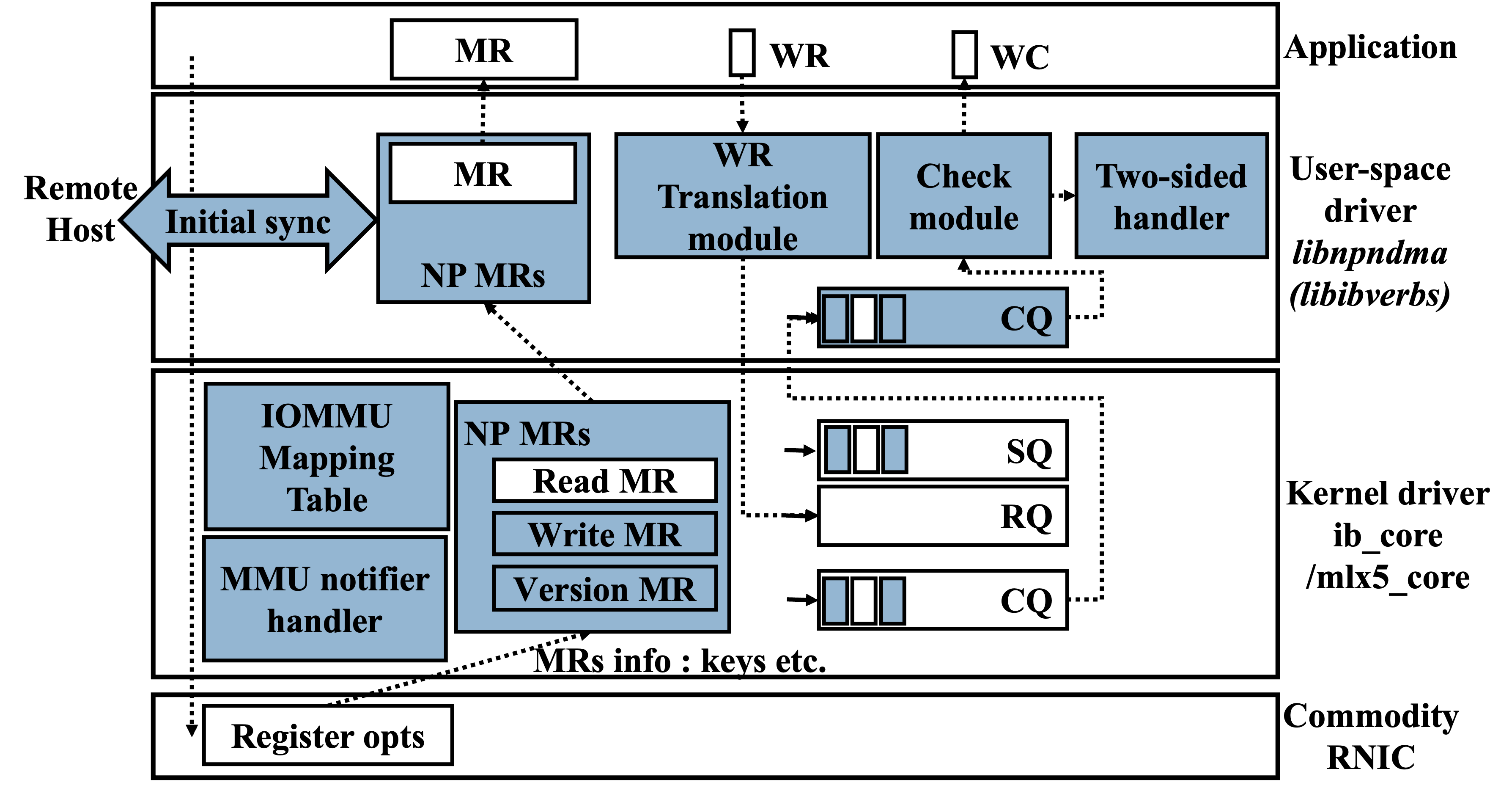}
	\vspace{-15pt}
	\caption{\label{fig:implementation}The implementation architecture of NP-RDMA.}
	\vspace{-15pt}
\end{figure}

NP-RDMA is implemented with an \texttt{LD\_PRELOAD} user-space library \texttt{libnprdma} with 5,500 LoC and a modified RDMA kernel module with a 2,800 LoC patch to Mellanox OFED 5.8~\cite{mellanox-ofed}.
It's architecture is shown in Figure \ref{fig:implementation}.

\subsection{Enabling Unmodified Applications}

To enable NP-RDMA for unmodified applications, we build an \texttt{LD\_PRELOAD} user-space library \texttt{libnprdma} which implements a subset of \texttt{libibverbs} APIs including MR, QP, CQ management on the control plane, and WQE posting, CQE polling and event waiting on the data plane.
The \texttt{libnprdma} APIs create MR and QP, CQ resources internally via a slightly modified user-kernel interface, so that Read/Write MRs can be differentiated by the kernel.
Other \texttt{libibverbs} APIs are kept unchanged, e.g., device and PD management.
To simplify implementation, NP-RDMA only supports RC mode and does not support experimental APIs, e.g., Mellanox direct verbs.

\textbf{Overwriting inline functions.}
A challenge is that some APIs, e.g., \texttt{ibv\_post\_send}, are inline functions in libibverbs, so we cannot replace it without re-compiling the applications.
To avoid re-compilation of applications, we observe that the inline functions actually invoke function pointers of the IB context.
So, we replace the internal post send functions when the context is created by \texttt{ibv\_open\_device}.

\textbf{Key mapping of auxiliary MRs.}
When the application creates an MR, three MRs are created: Read, Write, and Version.
The IOMMU mappings of the Read and Write MR are initialized according to the page mappings at registration time.
The version MR is a pinned MR where every 4 bytes represents a page, which is initialized according to whether the page exists.
Auxiliary Write and Version MRs introduce a challenge: MRs are referenced by \texttt{lkey} and \texttt{rkey}, which is allocated by both the driver and NIC.
The keys are exchanged by the application using side channel, but 3 keys cannot be packed into one key.
How can the initiator know the \texttt{rkey} of the auxiliary MRs given the \texttt{rkey} of Read MR?
To solve this problem, we synchronize the key mappings of auxiliary MRs between initiator and target whenever an MR or QP is created.
When a QP is first used to transmit a message, key mappings of existing MRs are sent through control commands.
This procedure introduces an extra RTT to exchange the remote key mappings.
Notice that QP creation is too early for synchronization because the other side may not be prepared.
After the initial synchronization, key mappings of newly created MRs are also sent through the QP.

\textbf{QP management.}
When the application creates a QP, only one QP is created, but the queue depth is 3x of the original.
This is because an application work request (WR) translates to at most 3 WRs in the NIC.
We also create a small pinned MR to transfer control commands, e.g., two-sided page fault handling.
The size of MR is 64 bytes multiplied by the QP depth.
For CQ, we create a user-space CQ and a NIC CQ for each application CQ.
A polling thread polls all NIC CQs in the process and generates a CQE to the user-space CQ when a WR is completed.
To support interrupt mode, NP-RDMA creates completion channels and \texttt{eventfd}s in user space.
When a CQ is in interrupt mode, the polling threads triggers the eventfd on WR completion, and the waiting application thread is waken up.

\textbf{Error handling.}
To achieve full compatibility, NP-RDMA needs to handle all error conditions, especially when control-plane operations (e.g., destroy QP or unregister MR) are concurrent with in-flight data plane operations.
We build a tool to generate hundreds of such combinations, and compare the behavior of standard IB verbs and NP-RDMA.

NP-RDMA also needs to free up resources when the process exits abnormally.
This is handled by the kernel module which monitors the close event of the IB device.

\subsection{IOMMU-based Page Mapping}

We do not make intrusive changes to the OS kernel.
All modifications are in the standard IB common modules and NIC drivers.
To enable IOMMU or SMMU, BIOS settings and kernel command line may need to change.
Applications not using NP-RDMA are not affected because the IOMMU performs identity mapping by default.
During initialization of the NP-RDMA driver, it allocates two pinned pages for signature and black-hole pages, which are shared globally.

\textbf{Updating page mapping on swap-out.}
During MR registration, rather than setting up the MTT to reflect the pinned page mapping, the driver configures the MTT to an immutable identity mapping, and then configures IOMMU/SMMU page table to map IOVAs to PAs by copying the OS page table.
Page-fault virtual addresses are mapped to the signature or black-hole page rather than NULL.
The driver also inserts MMU notifiers~\cite{mmu-notifier}, so the OS will synchronously invoke the callback when a page is swapped out in the MR range.
The callback updates the IOMMU/SMMU mapping of the corresponding Read/Write MRs to the signature and black-hole pages, and increments the version MR.
Then, it flushes the IOMMU/SMMU cache in case of in-flight DMA transactions.
Finally, the callback returns, and the OS is free to reallocate this page.

\textbf{Updating page mapping on swap-in.}
Although Linux supports swap-out event callbacks, it currently does not support swap-in.
So, we use a lazy approach where the first access to a swapped-in page is still considered a page fault in the optimistic checking, and two-sided page fault handling will update the IOMMU mapping.

\textbf{Handling race conditions.}
In two-sided page fault handling, the target temporarily pins the pages to guarantee successful one-sided Write/Read.
To avoid race conditions between concurrent pinning and unpinning, the NP-RDMA kernel module maintain a thread-safe hash table tracking the reference counts of all pinned pages.
A page is only unpinned when the reference count drops to zero.

\subsection{Supporting Non-Pinned Atomic/Send/Recv}
\label{subsec:design-send-recv}
Now we have discussed Read and Write verbs.
Atomic is different from Write because it is not idempotent.
The optimistic approach cannot distinguish between a coincidence with magic numbers and an actual page fault, so it is always retried with the two-sided approach.
However, if an atomic operation is executed twice, the result would be incorrect.
So, we always implement atomics with the two-sided approach and execute the atomics using host CPU on the target.

Send/Recv are easy to support.
The send and receive buffers can be temporarily pinned and mapped by IOMMU.
The initiator and target unpin the buffers when it receives the CQEs.
During production deployment, customers require further optimizations to the pinning of receive buffers, because receive buffers need to be posted into the RQ beforehand.
In these cases, we implement Send/Recv in a rendezvous manner, similar to the reverse read in two-sided page fault handling.
The initiator pins and sends the address of send buffer, then the target temporarily pins the receive buffer, performs reverse one-sided read, and unpins the receive buffer.

Write-with-immediate is implemented as a Write followed by a Send.
Upon reception of the Send command, the target checks magic numbers against the written data.
If the optimistic Write succeeds, it generates CQE and does not introduce extra RTT.
Otherwise, it notifies the initiator to use two-sided page fault handling.

\section{Performance Evaluation}
\label{sec:convergence-evaluation}
In this section, we evaluate the performance of NP-RDMA through a series of micro-benchmarks. 
First, \S\ref{subsec:non-page-fault} shows that \name incurs only slight overhead under normal non-page-fault operations. Second, \S\ref{subsec:page-fault} shows that \name can handle it fast when operations encounter page-faults. Finally, \S\ref{subsec:control-overhead} evaluates the states consumption and time cost in \name's control-plane.

\subsection{Environment Settings}
\label{subsec:testbed-setup}


\smalltitle{Testbed:} Our testbed consists of 2 machines (Dell R740, Intel(R) Xeon(R) Silver 4210R CPU @ 2.40GHz), with Mellanox CX-5 and CX-6 RNICs connected by 100G network cables. The driver version is Mellanox OFED 5.8, and the OS is Ubuntu 20.04 with Linux 5.11.1 kernel version. 


\smalltitle{Methods compared and the workload:} We compare \name with original RDMA and ODP. We use the existing tools in Perftest package~\cite{perftest} (v6.09) to test Read/Write latency and throughput, setting the affinity to using one CPU core. The results for each Read/Write message size is the average of 1000 runs. Our test is mainly aimed at RDMA one-sided operations, \ie, Read/Write. Send/Recv is originally designed to involve CPU which is easy to handle page fault as introduced in \S\ref{subsec:design-send-recv}, so it is not the focus of our evaluation.


\subsection{Non-page-fault Scenarios}
\label{subsec:non-page-fault}

\begin{figure*}[t]
\begin{minipage}[]{.34\textwidth}
	\centering
	\vspace{-10pt}
	\subfigure[Reads]{
		\label{fig:Lat-read-NPFs}
		\includegraphics[width=\textwidth]{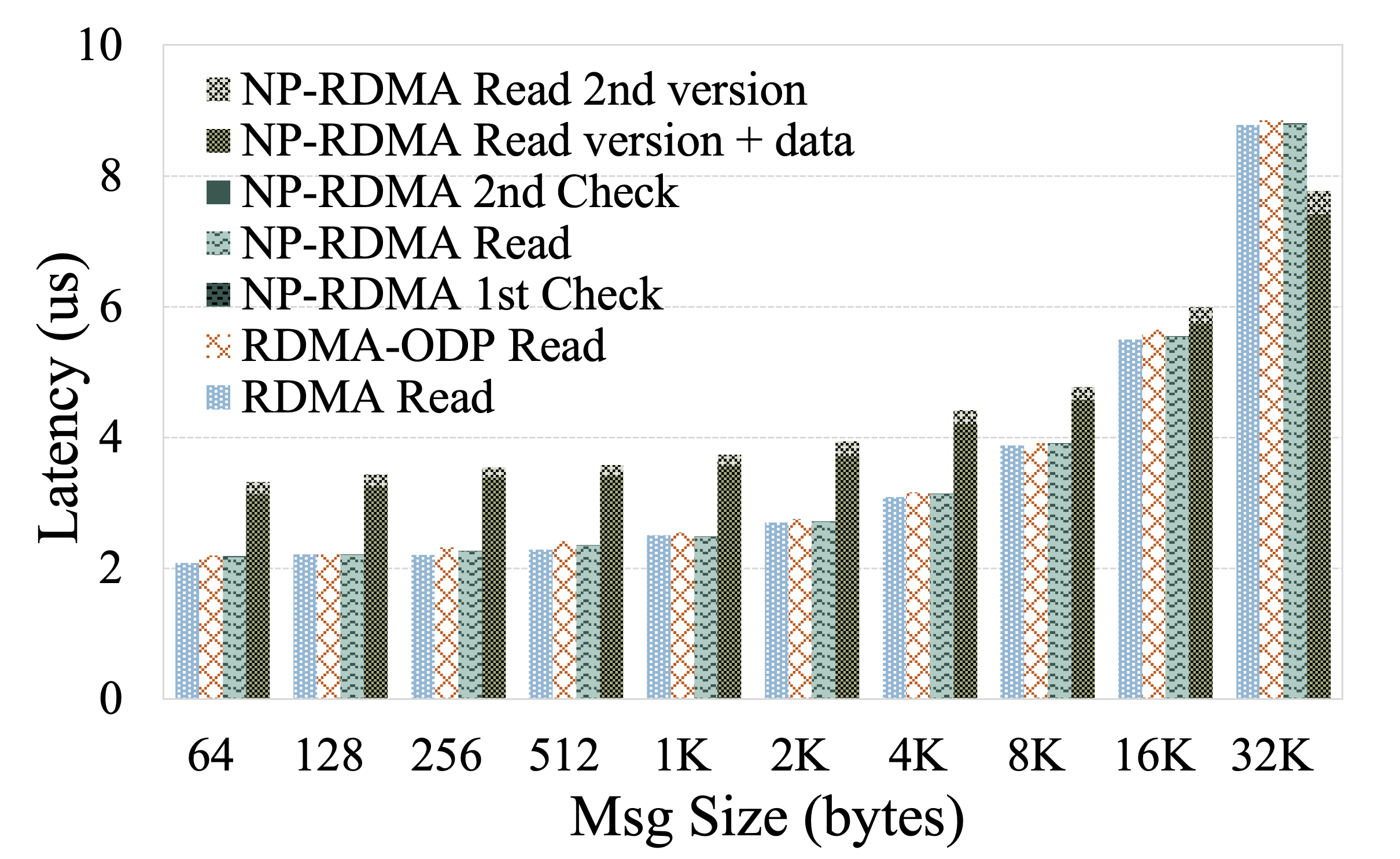}
	}
	\subfigure[Writes]{
		\label{fig:Lat-write-NPFs}
		\includegraphics[width=\textwidth]{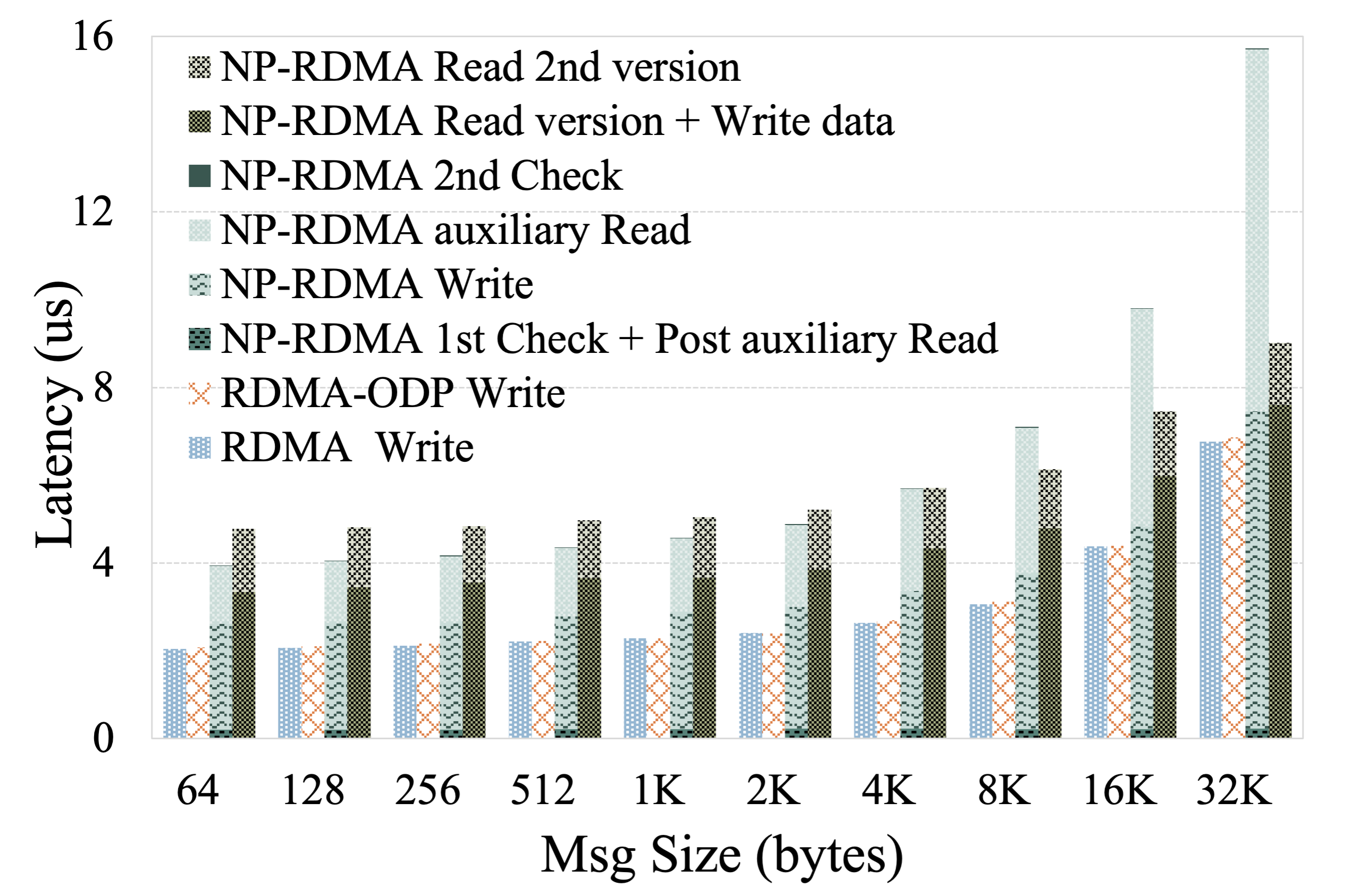}
	}
	\vspace{-15pt}
	\caption{Latency under no page faults.}
	\label{fig:latnpf}
	\vspace{-10pt}
\end{minipage}
\hspace{0.01\textwidth}
\begin{minipage}[]{.26\textwidth}
	\centering
	\subfigure[Read latency.]{
		\label{fig:pf-read-lat}
		\includegraphics[width=\textwidth]{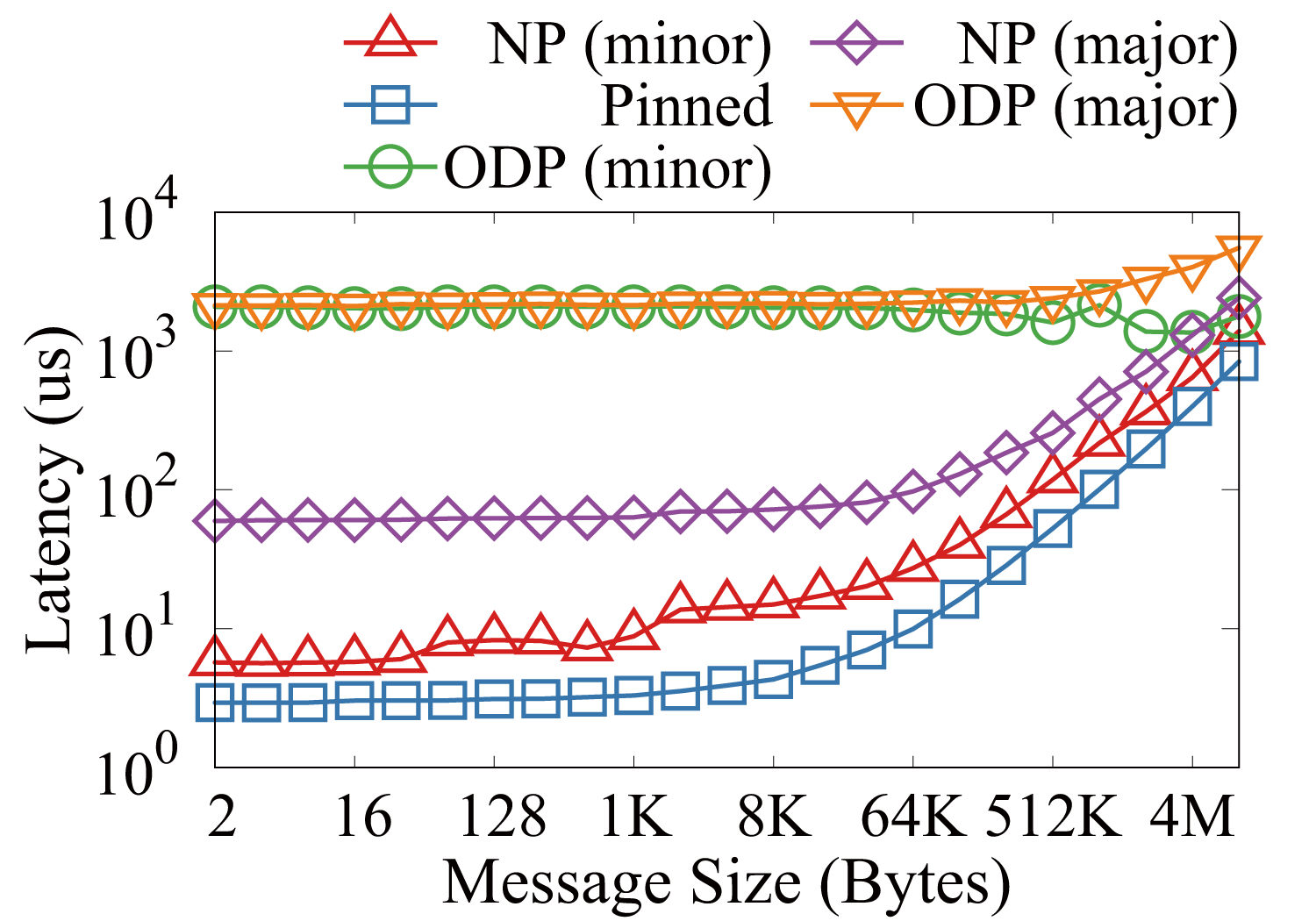}
	}
	\subfigure[Write latency.]{
		\label{fig:pf-write-lat}
		\includegraphics[width=\textwidth]{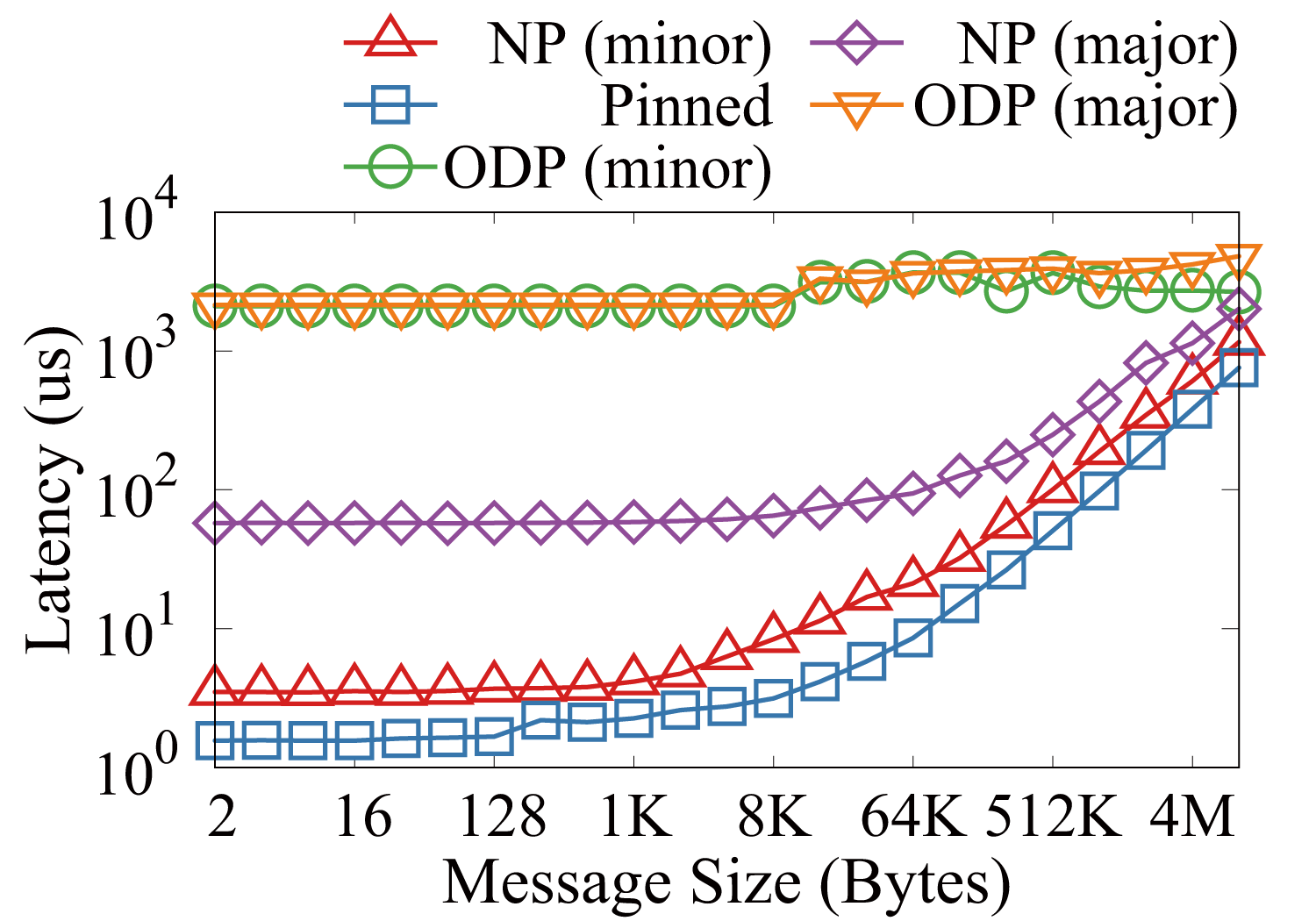}
	}
	\vspace{-15pt}
	\caption{Latency under page faults.}
	\label{fig:latency-page-fault}
	\vspace{-10pt}
\end{minipage}
\hspace{0.01\textwidth}
\begin{minipage}[]{.26\textwidth}
	\centering
	\subfigure[Read throughput.]{
		\label{fig:pf-read-tput}
		\includegraphics[width=\textwidth]{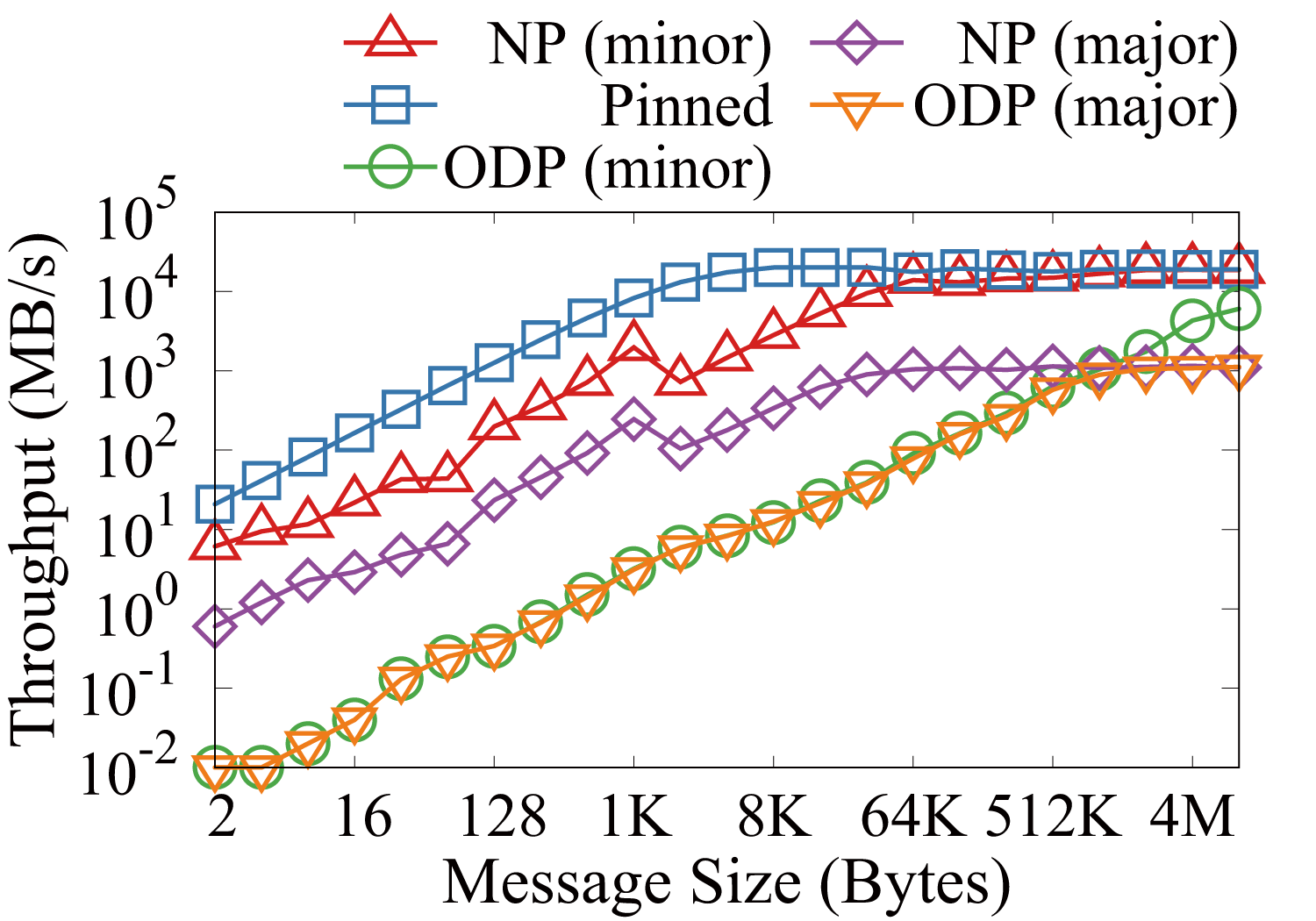}
	}
	\subfigure[Write throughput.]{
		\label{fig:pf-write-tput}
		\includegraphics[width=\textwidth]{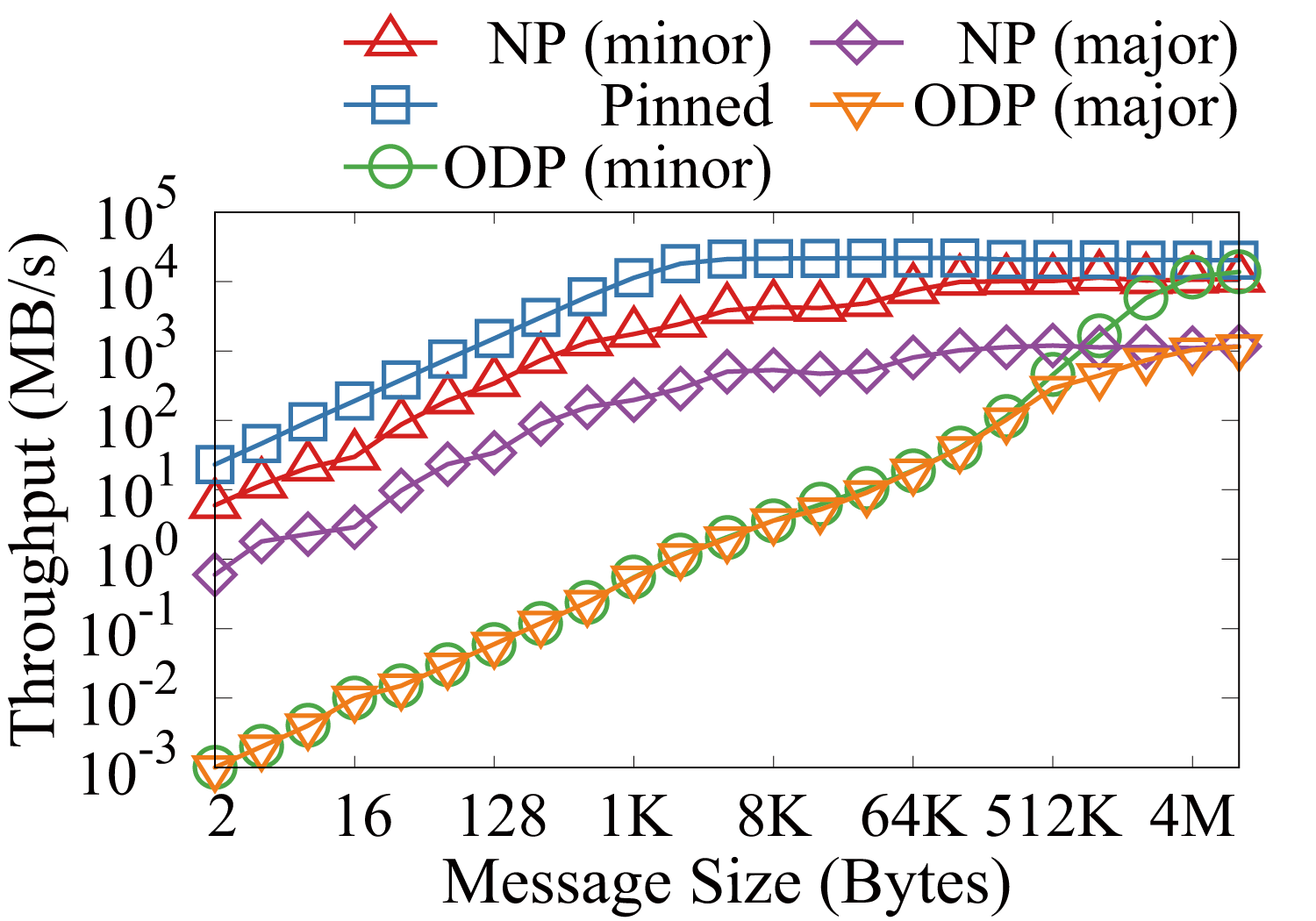}
	}
	\vspace{-15pt}
	\caption{Throughput under page faults.}
	\label{fig:performance-page-fault}
	\vspace{-10pt}
\end{minipage}
\end{figure*}

\smalltitle{Latency:} Figure \ref{fig:latnpf} shows the end-to-end latency and its breakdown for doing Read and Write operation with different message sizes. 
For Read operation (Fig.~\ref{fig:Lat-read-NPFs}), \name only incurs about 0.4\%$\sim$1\% extra time compared to original RDMA or ODP under various message sizes.
The signature page approach (denoted as \name 1st Check, \name Read, and \name 2nd Check) always outperform page versioning (denoted as \name Read version + data, \name Read 2nd version), because the CPU overhead for checking magic numbers is lower than issuing two additional WQEs.\footnote{In Fig.~\ref{fig:Lat-read-NPFs}, signature and page versioning approaches are tested with different NICs, so their latencies are not directly comparable.}

For Write operation that uses the signature page approach (Fig. \ref{fig:Lat-write-NPFs}), \name only needs about 0.5$\mu$s more to successfully write the actual data to the remote memory. 
With regard to the extra 0.5$\mu$s, 
only a small part ($\sim$0.2$\mu$s) comes from checking the data content and the initializing the auxiliary Read WR. Detailed examination shows the major delay is due to that processing the subsequent Read WR on RNICs may delay the former Write WR processing, although the former operation should not be affected by subsequent ones according to the RDMA specification. 
While, if we focus on the time when the initiator confirms that a Write is completed correctly, \name will take twice as long as RDMA. 
The overhead mainly comes from that, using the signature page approach (\S\ref{subsec:magic-number}), the auxiliary Read needs to read the same data back.
As such, we use this approach only for 4KB and shorter messages.
For larger Write messages, the page versioning approach outperforms signature page because it does not need to double the transfer size.



\begin{figure}[htbp]
    \centering
    \vspace{-10pt}
    \subfigure[Read]{
    \label{fig:bw-read-NPFs}
    \includegraphics[width=0.22\textwidth]{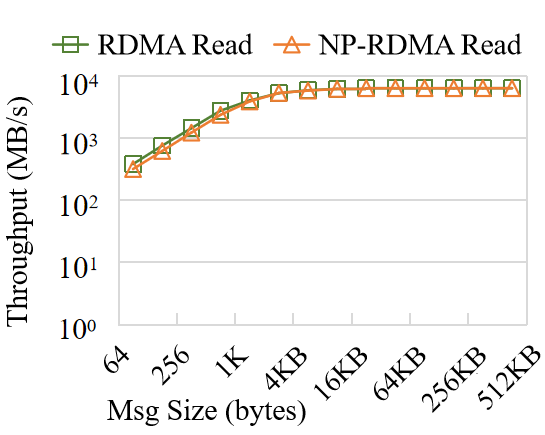}
    }
    \subfigure[Write (ib\_write\_bw)]{
    \label{fig:bw-write-NPFs}
    \includegraphics[width=0.22\textwidth]{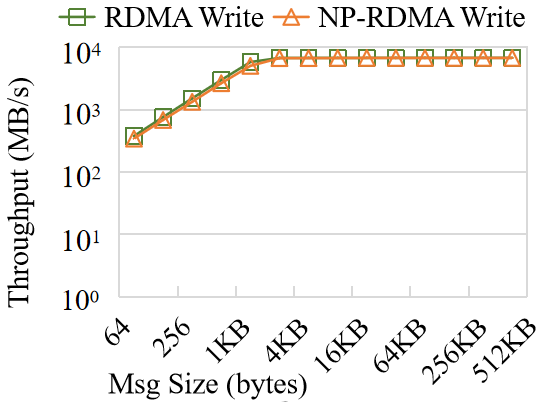}
    }
    \subfigure[Write (all signaled Writes)]{
    \label{fig:bw-write-signal}
    \includegraphics[width=0.25\textwidth]{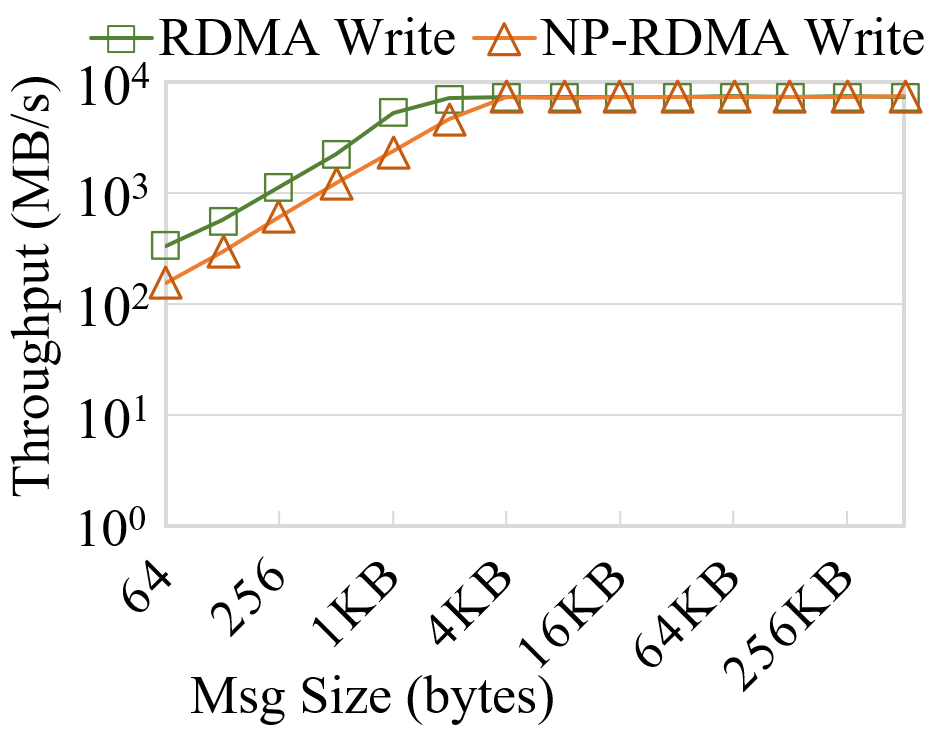}
    }
    \vspace{-15pt}
    \caption{Throughput under non-page-fault scenarios.}
    \label{fig:bwnpf}
    \vspace{-15pt}
\end{figure}

\smalltitle{Throughput:} We first use ib\_read\_bw/ib\_ write\_bw in perftest to evaluate the throughput of \name. Fig.~\ref{fig:bw-read-NPFs} and Fig.~\ref{fig:bw-write-NPFs} show that \name reach almost the same throughput as original RDMA. When testing Write bandwidth, Perftest periodically 100 unsignaled Writes followed by a signaled Write. Since \name batches the page-fault check of unsignaled Writes (\S\ref{subsec:magic-number}), it incurs only a slight CPU and throughput overhead. Such manner of using consecutive unsig\-naled Writes is common for most bandwidth-greedy applications that need to transmit large data. 

However, there are also some bandwidth-sensitive applications which may generate a large amount of signaled Writes (\eg, small write transactions in data bases). 
Fig.~\ref{fig:bw-write-signal} evaluates \name's performance under such scenarios, all using signaled Writes.  
Results show that for small Writes, \name's throughput is about half of original RDMA, because of processing an extra Read for each Write. However, as the message size grows up to 4KB, \name can saturate the 100Gbps bandwidth. 




\subsection{Page-fault Scenarios}
\label{subsec:page-fault}


To generate minor page faults, we allocate a large piece of memory with \texttt{mmap}.
Due to on-demand paging of Linux, remote accesses to it always generates minor page faults.
To generate major page faults, we first allocate large memory and initialize it, then run a memory pressure process that keeps allocating and accessing memory until the large memory is completely swapped out.

\smalltitle{Latency:} Figure \ref{fig:pf-read-lat} and Figure \ref{fig:pf-write-lat} show that \name only needs 3.5~$\mu$s (Read) or 5.7~$\mu$s (Write) to handle a minor page fault and complete the data transmission.
Compared with Mellanox ODP, \name reduces 2-byte message latency to 1/594 for read and 1/364 for Write, because ODP waits for a 2~ms timeout in Mellanox CX-5 NICs.
As shown in Sec.\ref{sec:background}, the Mellanox CX-6 NICs in our testbed have a 16~ms timeout.
For 2-byte messages, \name adds 2.8~$\mu$s (Read) or 1.9~$\mu$s (Write) to the latency of pinned RDMA due to its two-sided nature.
It does not introduce extra round-trips, though, because the data is sent inline.
For messages larger than 1KB, \name uses the reverse Write/Read mechanism to handle page faults, which introduces an extra round-trip and the need to temporarily pin the pages and update IOMMU mapping.
So, we see a jump of latency at the 2KB message size of Fig.~\ref{fig:pf-read-lat}, where the extra delay increases to 10~$\mu$s.
At 2~KB message size, \name still has 1/160 latency of ODP.
For larger messages, the overhead of \name is dominated by the time to resolve page faults.
For a 8~MB message, \name has 67\% higher latency than pinned RDMA.

For major page faults, \name only needs 60~$\mu$s (Read) or 57~$\mu$s (Write) which simply adds the time to swap in a page from SSD. 
The latency is 1 to half orders of magnitude lower than ODP.
For large messages, the latency is dominated by the throughput of SSD swap-in, which is roughly 1~GB/s on our testbed. 
The latency of a 8~MB message with major page fault is 1.7x of minor fault.

\smalltitle{Throughput:} 
Figure \ref{fig:pf-read-tput} and Figure \ref{fig:pf-write-tput} show that \name loses 3x Read throughput and 4x Write throughput for 2-byte messages when encountering minor page faults compared to pinned RDMA.
This is because a \name thread can only handle 1.5M minor page faults, which is the bottleneck of small message processing.
In contrast, a thread can issue 5M Read and 6M Write requests to the NIC using pinned RDMA.
Although the overhead of \name seems to be high, it is still 600x faster than ODP.

When \name converts from inline send mode to reverse Write/Read, the throughput sees a drop for Read, which is the worst case compared to pinned RDMA.
With 2~KB messages, \name loses up to 18x Read and 7x Write throughput due to the overheads in temporary pinning, IOMMU mapping update, reverse one-sided operations, and page fault handling.
For large messages, \name can almost keep the same Read performance as under non-page-fault scenarios and saturate the NIC bandwidth.
The Write throughput of \name saturates at half of NIC bandwidth due to limited credits in the flow control of NIC.
For large messages, e.g., 1~MB, \name still outperforms ODP by 19x (Read) and 7x (Write) because \name parallelizes page fault handling and thus can leverage the SSD throughput, while in ODP one page-fault message blocks all subsequent messages.


Because \name parallelizes page fault handling, it can leverage the SSD throughput better than ODP, where one page-fault message blocks all subsequent messages.

\subsection{Control-Plane Overhead}
\label{subsec:control-overhead}

\begin{table}[h]
	\centering
	\begin{tabular}{|c|c|c|}
		\hline
		\multirow{3}{*}{Per-page} & 4B & PTE of Write MR IOMMU mapping\\\cline{2-3}
		& 4B & PTE of Read auxiliary MR mapping\\\cline{2-3}
		& 4B & Version MR\\\hline
		\multirow{4}{*}{Per-QP} & 1MB & Pinned MR for control cmd queue\\\cline{2-3}
		& 128KB & Memory ranges of in-flight requests \\\cline{2-3}
		& 32KB & Atomic locks of temp pinned pages \\\cline{2-3}
		& 16$N_{MR}$ & Remote MR key mapping \\\hline
		Per-CQ & 128KB & User-space CQ \\\hline
	\end{tabular}
	\caption{States consumption in \name's control-plane}
	\label{tab:memory-overhead}
	\vspace{-15pt}
\end{table}

\begin{table}[h]
	\centering
	\begin{tabular}{|c|c|c|}
		\hline
		    & Original & NP-RDMA \\\hline
		    \hline
		Library init & 43 ms & 49 ms \\\hline
		Create MR & \parbox[t]{1in}{50 $\mu$s base +\\ 400 ms per GB} & \parbox[t]{1in}{135 $\mu$s base +\\ 20 ms per GB} \\\hline
		\hline
		Create QP & 45 $\mu$s & 67 $\mu$s \\\hline
		Create CQ & 29 $\mu$s & 56 $\mu$s \\\hline
		QP init & 12 $\mu$s & 19 $\mu$s \\\hline
		\hline
		Swap out & 75 $\mu$s & 78 $\mu$s \\\hline
	\end{tabular}
	\caption{Time cost in \name's control-plane.}
	\label{tab:control-plane-overhead}
	\vspace{-15pt}
\end{table}

Previous experiments show that \name incurs very slight overhead to data-plane operations, however, it do incurs some control-plane overhead which is not on the data path. 
The memory overhead to maintain auxiliary states of NP-RDMA is summarized in Table~\ref{tab:memory-overhead}. NP-RDMA introduces small time overheads on the control plane due to initialization of auxiliary data structures, key exchange during QP init, and creation of auxiliary MRs, as Table~\ref{tab:control-plane-overhead} shows.
Creating large MRs is much \textit{faster} by removing pinning.
Since IOMMU mappings need to be updated when a page is swapped out, the latency increases by 3~$\mu$s, but it is a small fraction of SSD latency.

\section{Applications and Deployment Experience}
\label{sec:application}

We deploy NP-RDMA in two production applications, namely Spark~\cite{zaharia2010spark} and enterprise storage.

\subsection{Spark}
\label{sec:eval-spark}
\begin{table}[h]
	\centering
	\begin{tabular}{|c|c|c|c|c|c|}
		\hline
		\multicolumn{2}{|c|}{Data Size} & 100GB & 200GB & 300GB & 1TB \\\hline
		\multirow{3}{*}{\begin{minipage}{0.04\textwidth}{Exec\\time}\end{minipage}} & Original & 725s & 1401s & 2138s & 7336s \\\cline{2-6}
		& NP-RDMA & 765s & 1425s & 2167s & 7340s \\\cline{2-6}
		& Slowdown & 5.4\% & 1.6\% & 1.3\% & 0.0\% \\\hline
		\multirow{4}{*}{\begin{minipage}{0.04\textwidth}{Mem\\(GB)}\end{minipage}} & Original & 14.4 & 26.0 & 36.0 & 68.8 \\\cline{2-6}
		& Swap-out & 12.4 & 20.8 & 30.6 & 46.1 \\\cline{2-6}
		& Physical & 2.0 & 5.2 & 5.4 & 22.7 \\\cline{2-6}
		& Savings & 86\% & 80\% & 85\% & 67\% \\\hline
	\end{tabular}
	\caption{Performance of Spark TPC-DS.}
	\label{tab:spark-performance}
	\vspace{-15pt}
\end{table}

The main challenge for Spark is to reduce initialization latency and use SSD as swap area when the system is under memory pressure.
Prior to using NP-RDMA, our Spark RDMA implementation pins the entire memory pool during initialization, which takes 120 seconds for 300 GB of memory pool.
With NP-RDMA, the initialization time reduces to 6 seconds.

By enabling swap area to store cold data and using on-demand paging, TPC-DS~\cite{nambiar2006making} applications on Spark reduce 67\% to 86\% physical memory utilization with only 0.0\% to 5.4\% performance slowdown, as shown in Table~\ref{tab:spark-performance}.

A challenge is that some deployment scenarios do not allow us to install kernel modules.
So, we design a pure user-space implementation for these scenarios.
Rather than using IOMMU to remap the pages, we always use the two-sided approach, and register standard MRs on the fly.
More concretely, we replace pinning and unpinning in steps 1), 3), 5), and 6) of Sec.\ref{subsec:two-sided} with MR registration and deregistration.
Since MR registration takes $\sim$50~$\mu$s, the latency of small work requests become 10$\sim$20x worse.
Fortunately, most Spark applications use RDMA to transfer large amounts of data, so the MR reg and dereg overhead is amortized.
No MR is actually registered during application MR registration, so the initialization time is even shorter: 4 seconds.
By enabling swap, TPC-DS application on Spark uses 82\% less physical memory with 10.9\% slowdown.

\subsection{Enterprise Storage}

\begin{figure}[h]
	\centering
	\includegraphics[width=2in]{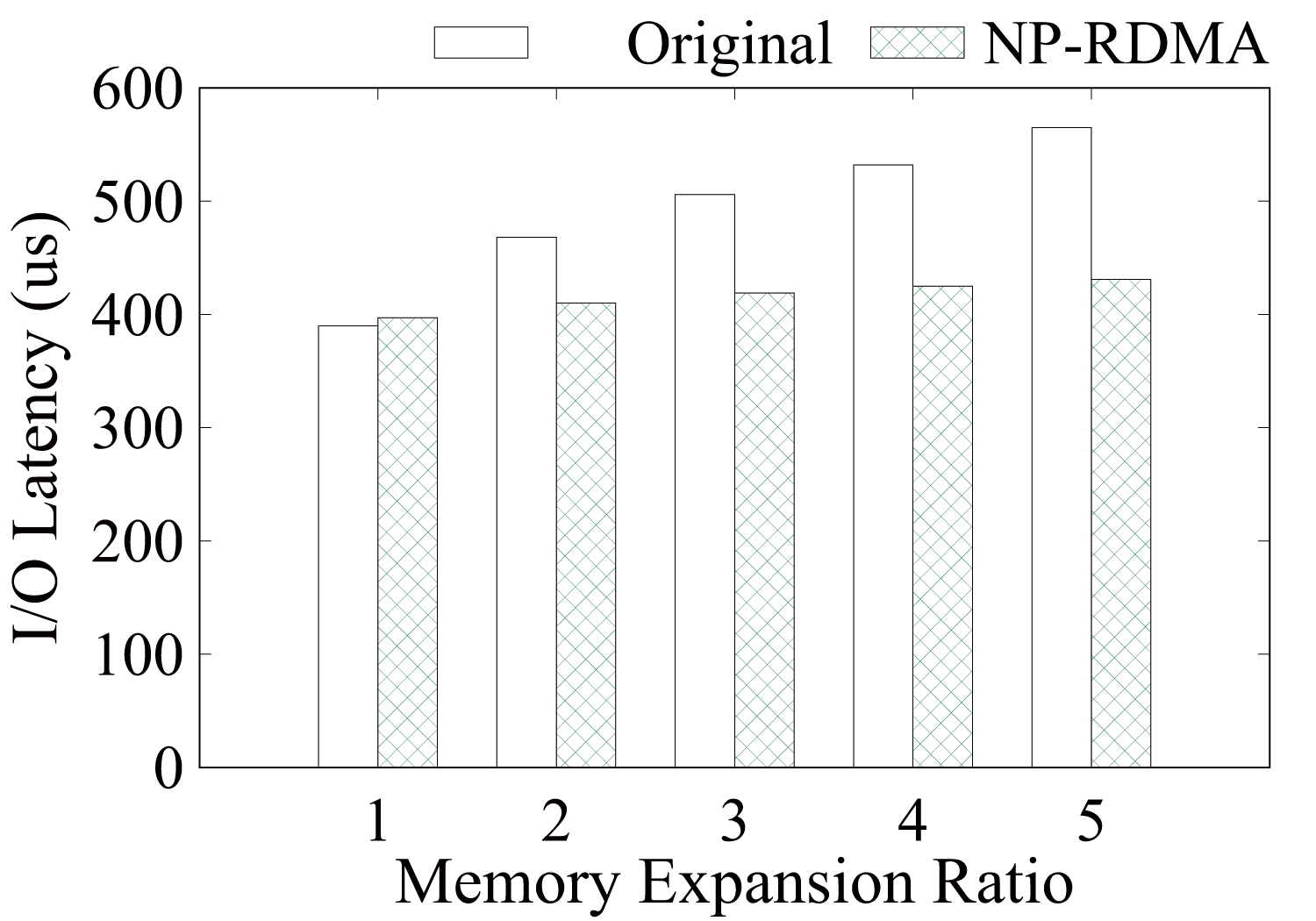}
	\vspace{-10pt}
	\caption{Performance of Enterprise Storage.}
	\label{fig:enterprise-storage}
	\vspace{-10pt}
\end{figure}

We enable NP-RDMA for an in-house enterprise storage application.
Back-end servers in the storage cluster maintains a memory pool as caching layer for persistent storage, which needs to be remotely accessed by the front-end servers of the storage cluster.
Traditionally, to avoid pinning the entire memory pool, the storage application maintains pinned send and receive buffers, and copies data from and to the application buffers.
With NP-RDMA, front-end servers can enjoy the low latency, high throughput and low CPU overhead of one-sided RDMA for cache-hit accesses.
We use IO-500~\cite{kunkel2016establishing} to generate the workload.
Figure~\ref{fig:enterprise-storage} shows the end-to-end IO latency of the enterprise storage.
Compared to the traditional approach, the average 8KB IO latency of NP-RDMA decreases by up to 24\% because of the removal of remote CPU involvement in cache-hit scenarios.
Compared to a pure in-memory baseline without any persistent storage, NP-RDMA increases average latency by 10\%, which mostly attributes to the SSD access latency of cache-miss Read/Writes.

In production deployment, the enterprise storage develops a customized memory swapping mechanism.
It allocates large chunks of physical memory during initialization, and maintain MMU mappings internally.
So, we can no longer rely on the OS MMU notifiers to update IOMMU/SMMU mappings.
Instead, we provide two callbacks for the customized memory swapping mechanism to notify swap-in and swap-out events.

Enterprise storage also has strict security requirements that disallow reverse one-sided Write/Read in two-sided page fault handling.
We find that after the page fault is resolved, the one-sided Read/Write can be simply retried using the optimistic one-sided approach.
So, the target sends a Receiver Ready message to the initiator, and the initiator retries, which only introduces 0.1$\sim$2~$\mu$s extra latency due to optimistic checks.

\section{Discussion}

\smalltitle{Limitations of NP-RDMA.}
The latency, bandwidth, CPU and memory overheads of NP-RDMA has been evaluated in Sec. \ref{sec:convergence-evaluation}.
During execution of Read and Write operations, the buffers may be written twice when the optimistic transfer actually succeeds but considered as failed due to concurrency or coincidence with magic numbers.
Although commodity NICs guarantees each address is only written once, writing twice is allowed in the Infiniband specification~\cite{infiniband-spec} due to idempotence of Reads and Writes.

Commodity NICs do not support hardware RDMA virtualization~\cite{kim2019freeflow,he2020masq}, so using IOMMU/SMMU for NP-RDMA would not affect NICs' virtualization features.

\smalltitle{Implications for hardware NP-RDMA.}
A NIC that can gracefully handle page faults may be more efficient than our software NP-RDMA approach.
However, some modifications should be made to the current RDMA abstractions and protocol.
First, to avoid head-of-line blocking caused by page faults, the NIC should provide out-of-order execution ability in a QP by enabling the applications to specify ordering constraints of work requests (Sec.\ref{subsec:ordering}).
Second, after the target software resolves a page fault, the operation must be resumed.
One approach is to buffer the failed operation at the target.
This would require significant scratchpad memory for Write, which is often not available on the NIC.
The other approach is to notify the initiator to retransmit the message.
This requires addition of a new notification message type to the RDMA protocol.

Hardware NP-RDMA still requires page table synchronization between OS and NIC.
First, currently Linux only supports MMU notifiers on swap-out, so swap-in events cannot be monitored.
We expect future work to improve it.
Second, updating NIC MTTs dynamically as in Mellanox ODP is a feasible solution, but it requires communication over PCIe on page swap.
We propose to use IOMMU/SMMU to perform virtual address mapping rather than MTTs, reducing synchronization overhead.
An ultimate solution we expect in future architectures is to enable IOMMU/SMMU to use MMU page tables directly, so that the software no longer need to maintain and synchronize two page tables.
\section{Conclusion}

This paper presents NP-RDMA, the first efficient software approach to using commodity RDMA without pinning memory.
Its key innovation is an optimistic one-sided approach that maps page-fault virtual addresses to special signature and black-hole pages using IOMMU/SMMU, which enables the initiator to detect page faults without actually triggering page faults on the target NIC.
NP-RDMA achieves close-to-zero overhead for non-page-fault accesses and only add two extra RTTs for page-fault accesses.
With non-pinned memory, Spark and enterprise storage applications increases memory capacity to 5$\sim$7x with less than 10\% performance drop.

\textit{This work does not raise any ethical issues.}


\end{document}